\documentclass[12pt]{article}
\usepackage{graphicx}
\usepackage[margin=1in]{geometry}
\usepackage{setspace}		
\usepackage{xcolor}
\usepackage{soul}
\usepackage{amsmath}
\usepackage{amssymb}
\usepackage{comment}
\usepackage{algorithm}
\usepackage{algpseudocode}
\usepackage[hidelinks]{hyperref}
\usepackage{graphicx}
\usepackage{mathtools}
\usepackage{multirow}
\usepackage{dsfont}
\usepackage{wrapfig}
\usepackage{titling}
\usepackage{setspace}
\usepackage{sectsty}
\usepackage{natbib}
\usepackage[labelfont=bf]{caption}
\usepackage{amsfonts}
\DeclareMathSymbol{\shortminus}{\mathbin}{AMSa}{"39}

\algnewcommand\algorithmicforeach{\textbf{for each}}
\algdef{S}[FOR]{ForEach}[1]{\algorithmicforeach\ #1\ \algorithmicdo}

\newcommand\simiid{\stackrel{\mathclap{\normalfont\mbox{\small{iid}}}}{\sim}}

\renewcommand{\thesection}{\arabic{section}.}

\sectionfont{\normalfont\fontsize{14}{12}\bfseries}
\subsectionfont{\normalfont\fontsize{14}{12}\selectfont}

\begin{document}

\begin{center}
\LARGE
    Direct-Assisted Bayesian Unit-level Modeling for Small Area Estimation of Rare Event Prevalence
\end{center}

\hspace{10pt}

\large
\noindent Alana McGovern$^{1*}$, Katherine Wilson$^2$, and Jon Wakefield$^{1,2}$

\hspace{8pt}

\small  
\noindent$^1$ Department of Statistics, University of Washington, Seattle WA, USA\\
$^2$ Department of Biostatistics, University of Washington, Seattle WA, USA

\hspace{8pt}
\normalsize

\noindent $^*$Corresponding author, amcgov@uw.edu

\newpage
\doublespacing
\begin{abstract}
    \noindent Small area estimation using survey data can be achieved by using either a design-based or a model-based inferential approach. Design-based direct estimators are generally preferable because of their consistency, asymptotic normality, and reliance on fewer assumptions. However, when data are sparse at the desired area level, as is often the case when measuring rare events, these direct estimators can have extremely large uncertainty, making a model-based approach preferable. A model-based approach with a random spatial effect borrows information from surrounding areas at the cost of inducing shrinkage. As a result, estimates may be over-smoothed and inconsistent with design-based estimates at higher area levels when aggregated. We propose two unit-level Bayesian models for small area estimation of rare event prevalence which use design-based direct estimates at a higher area level to increase consistency in aggregation. This model framework is designed to accommodate sparse data obtained from two-stage stratified cluster sampling, which is particularly relevant to applications in low- and middle-income countries.  After introducing the model framework and its implementation, we conduct a simulation study to evaluate its properties and apply it to the estimation of the neonatal mortality rate in Zambia, using 2014 Demographic Health Surveys data. 
    \newline

    \newline
    
    \noindent \textbf{Statement of significance}: This work contributes to the literature regarding small area estimation of rare event prevalence using sparse complex survey data - such applications are extremely important in low and middle income countries in particular.  We introduce a Bayesian model which integrates design-based estimators into a unit-level hierarchical model, which encourages agreement between aggregated small area estimates and design-based estimates at a higher area level, which will be advantageous in global and public health settings.
\end{abstract}

\raggedright
\setlength{\parindent}{20pt}

\section{\MakeUppercase{Introduction}}

In terms of simplicity and fewest assumptions, the ideal method for obtaining estimates from complex surveys is a design-based weighted estimator, such as the Horvitz-Thompson (\citealp{HTestimator}) or H{\'a}jek estimator (\citealp{hajek}). These estimates use sampling weights which incorporate the sampling probability, and possibly non-response and post-stratification adjustments. While these design-based weighted estimators are consistent and asymptotically normal (\citealp{asymptotics}), they also produce very large design-based variance estimates when data are sparse. In small area estimation (SAE) problems there is often insufficient data to produce design-based estimates with reasonable precision, making it necessary to use a model-based approach with random effects. By borrowing information across areas, precision is increased at the cost of introducing some shrinkage bias (\citealp{Datta2012,Wakefield2020}). The bias introduced by these model-based approaches often makes higher level combinations of these small area estimates inconsistent in aggregate. For example, the aggregation of state-level model-based estimates may not be consistent with the national design-based estimate.

There have been many procedures introduced in the SAE literature which ensure aggregation accuracy of small area estimates to more reliable, higher-level estimates, referred to as benchmarks. There is a vast literature on this topic and we highlight a relevant subset here. One early development is the use of nested error linear regression models for estimating small area means (\citealp{pfeffermann1991,YouRao2002}).  \cite{Wang2008} derived a unique best linear unbiased estimator for small area means from augmented area-level linear models under a single benchmarking constraint, and \cite{BellDattaGhosh2013} extended this result to accommodate multiple benchmarking constraints. \cite{Datta2011} developed a class of benchmarked Bayes estimators under area-level generalized linear models.  All of these methods modify the best linear unbiased predictor of the small area means to achieve the benchmarking constraint, which leads to increased variance under the model (\citealp{BellDattaGhosh2013}). \cite{BergFuller2018} proposed two benchmarking procedures for nonlinear models: one which uses a linear additive adjustment, and the other which uses an augmented model for the expectation function.  All of the aforementioned methods consider benchmark constraints which are estimated from the same data. There is also a body of work exploring the use of posterior projections to conduct Bayesian inference in constrained parameter spaces (\citealp{Dunson}; \citealp{Sen}; \citealp{Nicolo}).  As \cite{Sen} notes, this framework is quite sensitive to misspecification of constraints, which implies that it would not provide reliable estimates when using design-based weighted estimates as constraints because of their large uncertainty. \cite{ZhangBryant} and \cite{NandramSayit} proposed inexact Bayesian benchmarking methods which can incorporate the uncertainty of the benchmarks into the posterior distribution via a soft constraint. However, both of these frameworks require that the benchmarks and small area estimates are produced from separate data sources. Violating this assumption would yield benchmarked small area estimates that underestimate uncertainty (\citealp{OkonekWakefield}).

The contribution of this paper is a pair of unit-level Bayesian mixed effects models with a likelihood modified to incorporate higher-level design-based direct estimates obtained from the same data source.  The set of models we will introduce use design-based estimates as constraints while accounting for the fact that these constraints are estimated from the same data. As \cite{StefanHidiroglou2021} note, ``statistical agencies favor an overall agreement between the sum of the model-based small area estimates and the direct estimate at a higher level that corresponds to the union of the small areas". The two models differ in that one enforces a hard constraint, while the other uses a soft constraint, accounting for the uncertainty of the benchmark. The goal of the latter is fundamentally different from exact benchmarking procedures, because its primary goal is to use higher-level estimates to encourage consistency in aggregation while taking the uncertainty of the benchmark into account, as opposed to imposing a hard benchmark constraint. Consistency in aggregation with design-based estimates is not only desirable for logistical reasons, but also because these estimates take the survey design into account, which model-based estimates typically do not. We propose two variations of our model because while taking the uncertainty of the benchmark into account is the more principled approach, there may be practical cases in which exact benchmarking is preferred by the practitioner.  We will focus on the estimation of rare event prevalence, as this is a setting in which issues of data sparsity tend to be most severe, making this direct-assisted unit-level model necessary. Examples of commonly measured outcomes that can be considered rare events, depending on the study population, include neonatal mortality, vaccination (or no vaccination) status, and HIV status. While the monitoring of these rare events at subnational levels is of substantive importance, the exceedingly small number of sampled events poses unique challenges in prevalence estimation.  

In section 2 we will introduce the context of SAE in low- and middle-income countries (LMICs), which motivates the new model. In section 3 we will present a standard Bayesian unit-level model for rare events and in section 4 we will extend this model and introduce the soft and hard direct-assisted Bayesian unit-level (DABUL) models and discuss its implementation. In section 5 we conduct a simulation study comparing the DABUL models to a standard unit-level Bayesian model. Section 6 presents an application of the DABUL models to estimation of the neonatal mortality rate (NMR) in Zambia and a corresponding cross-validation study. We conclude with a discussion in Section 7.

\section{\MakeUppercase{Small area estimation in low- and middle-income countries}}

It is common to utilize nationally representative samples collected by the Demographic and Health Surveys (DHS) (\citealp{dhs}) and Multiple Indicator Cluster Surveys (MICS) (\citealp{mics}) to obtain estimates in LMICs at the level of the first or second administrative area.  DHS and/or MICS carry out surveys in the majority of LMICs, often as frequently as every five years.  The DHS and MICS are conducted using a two-stage stratified cluster sampling design, where the sampling strata are defined by urban/rural crossed with the first (or for some countries, second) administrative area. In the first stage, a selection of enumeration areas (EAs) are sampled from each strata, where the probability a given EA will be selected is proportional to the number of households in that EA, relative to the others in its strata. In the second stage, a fixed number of households are sampled with equal probability from each EA. Under this design, a `cluster' refers to either an EA or a segment of an EA.

The NMR is one of a multitude of demographic and health outcomes that is often estimated from these surveys. Accurate and precise estimation of the NMR at subnational levels is particularly important because it allows government officials at the country or regional level to evaluate which regions need more targeted interventions. The Sustainable Development Goals (SDGs) provide a target of no more than 12 deaths per 1000 live births by 2030 (\url{https://sdgs.un.org/2030agenda}).  As previously stated, using a design-based weighted estimator to estimate the subnational NMR would be ideal, but is often not feasible due to small area-level sample sizes and the relative rarity of neonatal death. The sampling frame for most DHS and MICS are powered at the first administrative level, so estimates at smaller area levels using design-based methods, or even area-level models, such as the Fay-Herriot model, are often not reliable. However, estimates at the second administrative level are desired because this is often the level at which health interventions are administered.  As a result of this sampling design, it is often helpful to use unit-level models to obtain small area estimates (\citealp{rao2015}), where the units are the sampled clusters.  Because the households in each cluster are recorded with the same location and sampling weight, it is natural to combine data within the same cluster and model these summed counts.

The methods of this paper are motivated by estimation of the NMR in cases where design-based estimates are reliable at the first, but not second, administrative level. The model can easily be simplified to accommodate a case where the design-based estimates are reliable at the national level, but not the first administrative level.  We will use NMR-specific terminology throughout (i.e., births and neonatal deaths), but our proposed model can be applied to prevalence estimation of any rare event using complex survey data. For example, if HIV status is the indicator of interest, `births' can be replaced by `individuals' and `neonatal deaths' replaced by `individuals with positive HIV status'.

\section{\MakeUppercase{Bayesian unit-level models for rare event prevalence}}

\subsection{Sampling Model}

Suppose we seek to estimate the prevalence of a rare event, neonatal mortality, at the second administrative level using sparse survey data. It is common to fit such a model under the assumption that the number of events follows an overdispersed Poisson distribution (\citealp{Diggle}), where the overdispersion accounts for the within-cluster dependence in outcomes. Specifically, let $n_c$ and $Z_c$ represent the number of births and neonatal deaths in a fixed time period in the sampled households in sampled cluster $c$, respectively, where each cluster is contained in one of $m_2$ second administrative areas. Then, for neonatal mortality rate in cluster $c$ (i.e., probability of death in the neonatal population in cluster $c$), $r_c$, and overdispersion parameter $\lambda$, we assume that for each cluster, $c$,

\begin{equation}
    Z_c|r_c,\lambda \sim \mbox{Negative-binomial}(n_cr_c,\lambda)
    \label{eq:negbin}
\end{equation}

\noindent with the parameterization

\begin{equation}
    P(Z_c|n_cr_c,\lambda ) = \frac{\Gamma(Z_c+n_cr_c/\lambda )}{\Gamma( Z_c+1)\Gamma(n_cr_c/\lambda )}\lambda^{Z_c}\left(1+\lambda\right)^{-(Z_c+n_cr_c/\lambda )}.
    \label{eq:negbinparam}
\end{equation}

Under this parameterization, $\mathbb{E}[Z_c|n_c,r_c,\lambda] = n_cr_c$ and Var$(Z_c|n_c,r_c,\lambda ) = (1+\lambda )n_cr_c$. We link this distribution with the regression model, 

\begin{equation}
\log(r_c) =\alpha + b_j\mathds{1}_{c \in \delta_2(j)}
\label{eq:basic}
\end{equation}

\noindent where $b_j$ is a second administrative level random effect (independent and identically distributed or with spatial structure) and $\delta_k(j)$ is the set of clusters in $k^{th}$ administrative area $j$. \ref{notation appendix} provides a summary of definitions and notation introduced throughout this paper. When this model is fit with sparse data there is often a large amount of shrinkage because information in each second administrative area is very limited. One way to mitigate this shrinkage effect is to use a regression model with nested spatial effects which includes a set of fixed effects at the first administrative level and a set of random effects at the second administrative level, i.e., replacing (\ref{eq:basic}) with

\begin{equation}
\log(r_c) =\alpha + \beta_i\mathds{1}_{c \in \delta_1(i)} + b_j\mathds{1}_{c \in \delta_2(j)}
\label{eq:basicnested}
\end{equation}

\noindent where $\beta_i$ is a first administrative level fixed effect and $\beta_1$ is fixed at $0$ to preserve identifiability.

\subsection{Motivating example: NMR in Zambia}

\begin{figure}[h]
    \centering
    \caption{{\bf Summary of births and neonatal deaths between 2009 and 2013 recorded in the 2014 Zambia DHS.} The top row provides maps of the total number of observed births in each first and second administrative area. The bottom row provides maps of the total number of neonatal deaths in each first and second administrative area. Each map has a different color scale to make differences between areas more clear.}
    \includegraphics[scale=0.3]{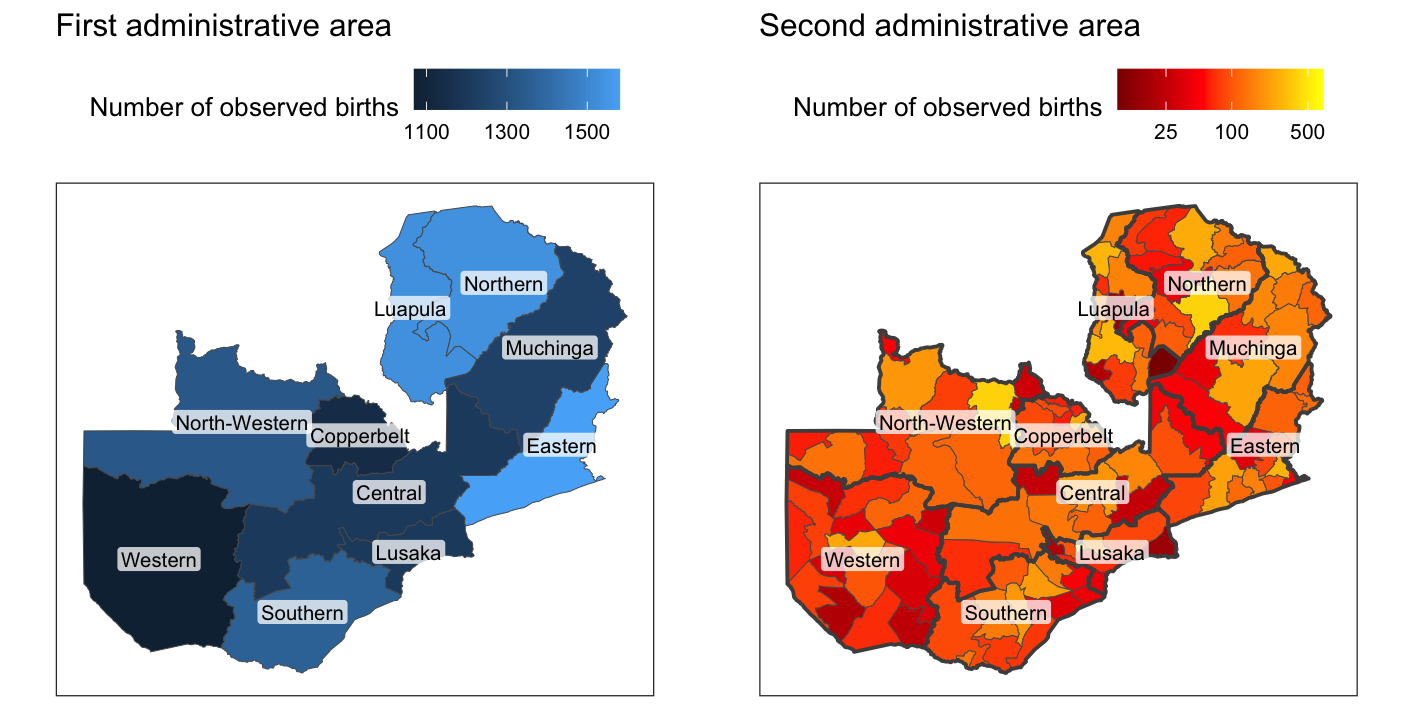}
    \includegraphics[scale=0.28]{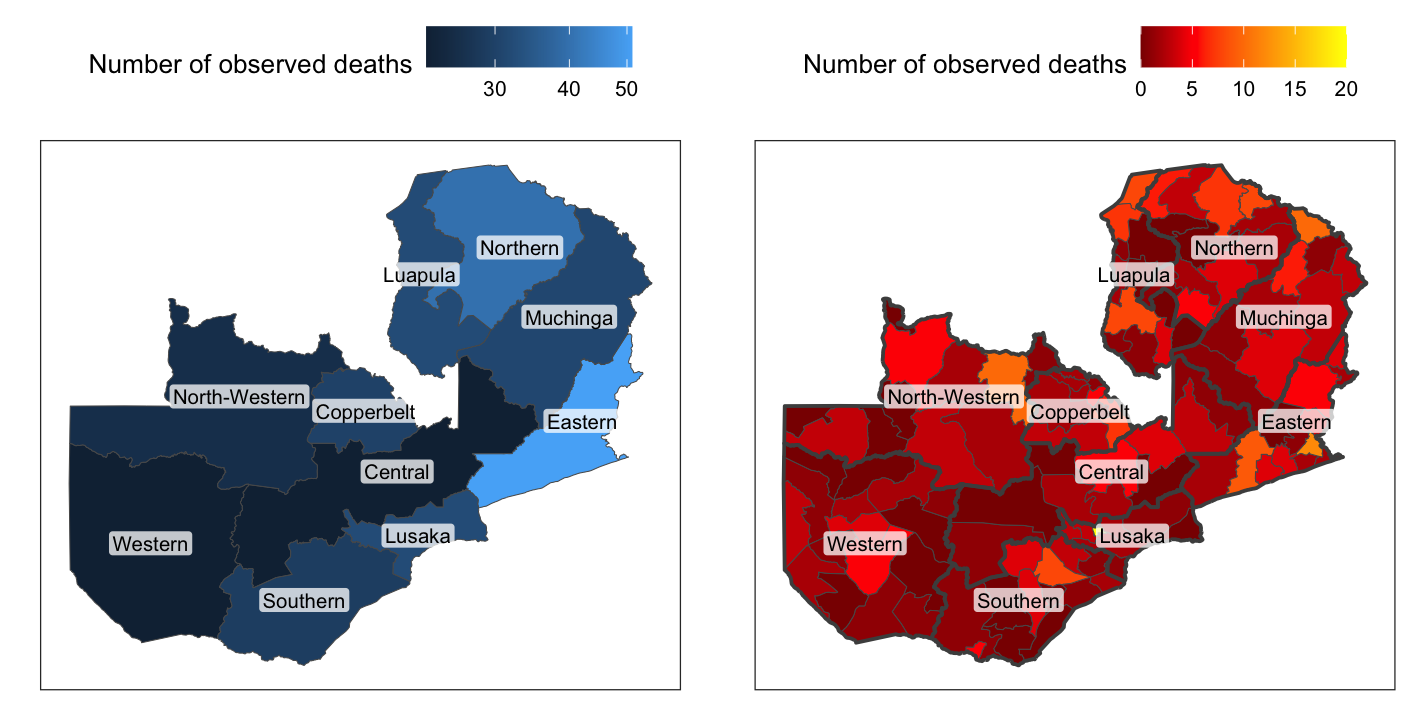}
    \label{fig:ZambiaData}
\end{figure}

To demonstrate the difference between regression models (\ref{eq:basic}) and (\ref{eq:basicnested}), consider the example of NMR estimation in Zambia at the second administrative level using all births between 2009 and 2013 sampled in the 2014 Zambia DHS (\citealp{dhsdata}). Zambia has 10 first administrative areas and 115 second administrative areas. A summary of these data is displayed in figure \ref{fig:ZambiaData}. The numbers of observed births and neonatal deaths in each first administrative area are sufficiently large to use an area-level model, such as Fay-Herriot (\citealp{FHmodel}), with an average of 1319 births and 32 deaths observed in each area. However, the number of observed births and neonatal deaths in each second administrative area are prohibitively small, with an average of 115 births and 2.8 deaths observed in each area, and 21.7\% of areas observing no neonatal deaths at all. At this level of data sparsity, a unit level model is useful for estimation at the second administrative level (\citealp{twocultures}). 

We must account for the stratified sampling design of the DHS survey (first administrative area crossed with urban/rural), so instead of estimating a global intercept, we include one urban and one rural intercept as follows,

\begin{equation}
\log(r_c) =\alpha_U\mathds{1}_{c \in \delta_U} + \alpha_R\mathds{1}_{c \in \delta_R} + b_j\mathds{1}_{c \in \delta_2(j)}
\label{eq:strat}
\end{equation}

\begin{equation}
\log(r_c) =\alpha_U\mathds{1}_{c \in \delta_U} + \alpha_R\mathds{1}_{c \in \delta_R} + \beta_i\mathds{1}_{c \in \delta_1(i)} + b_j\mathds{1}_{c \in \delta_2(j)},
\label{eq:stratnested}
\end{equation}

\noindent where $\delta_U$ and $\delta_R$ are the sets of urban and rural clusters, respectively, and $\beta_1$ is fixed at $0$ to preserve identifiability.

We estimate the NMR at the second administrative level under regression models (\ref{eq:strat}) and (\ref{eq:stratnested}) using the \texttt{Stan} software (\citealp{standev}). While a variety of random effect models at the second administrative level could be employed, we use a BYM2 spatial effect. Introduced by \cite{BYM2}, the BYM2 model is a re-parameterized version of the Besag-York-Molli\'e (BYM) model (\citealp{BYM}), which includes both unstructured, independent and identically distributed (IID), spatial effects and structured, intrinsically conditional autoregressive (ICAR), spatial effects, (\citealp{ICAR}). The BYM2 model has two parameters: $\sigma^2$, which indicates the total variance of the random effects, and $\phi$, which indicates the proportion of this variation that is explained by the structured component. The structured component has a sum-to-zero constraint to ensure identifiability. More specifically, in (\ref{eq:stratnested}), the structured component of the spatial effect has a separate sum-to-zero constraint for the areas within each first adimistrative area, while in (\ref{eq:strat}) there is only one global sum-to-zero constraint. In \ref{spatial_models} we compare NMR estimates and uncertainty under model (\ref{eq:stratnested}) using IID and BYM2 spatial effects, which shows that, in this example, choice of spatial effect model makes little difference, so we choose BYM2 as it provides the option of more structure (by including an ICAR component in addition to an IID component) and has the added benefit of allowing more spatially informative estimation in areas without observed data. For hyperpriors, we set $\phi\sim Beta(0.5,0.5)$ and use a penalized complexity (PC) prior for $\sigma^2$ with hyperparameters $U=1$ and $\alpha=0.01$, which corresponds to the prior belief that $P(\sigma>1) = 0.01$ (\citealp{PCPriors}). We place diffuse priors on the overdispersion and regression parameters: $\lambda\sim \mbox{Exp}(1)$, $(\alpha_U,\alpha_R)\simiid\mathcal{N}(-3.5,3^2)$ and $\beta_i\sim\mathcal{N}(0,1)$. The priors on the urban and rural intercepts are centered at -3.5 to reflect the prior belief that neonatal death is a rare event ($e^{-3.5}\approx 0.03$). Neonatal mortality rate estimates in many Sub-Saharan African countries, including Zambia, which are available on the UN-IGME Child Mortality Estimation dashboard (\texttt{childmortality.org}), confirm that this is a reasonable prior mean for the overall level. The priors on the regression parameters $(\alpha_U,\alpha_R,\boldsymbol{\beta})$ are chosen with the goal of being relatively diffuse on the exponential scale, while still accounting for the prior belief that the
outcome measure is a rare event.  More information regarding these prior choices is presented in \ref{Priors Appendix}. We aggregate the urban and rural estimates within each second administrative area using urban/rural population fractions estimated using the method in \cite{wu}. This method is a logistic classification model which uses pixel-level population density and defines a threshold for classifying pixels as urban or rural. This threshold is calibrated by urban/rural composition at the first administrative level which is reported by DHS (\citealp{dhs}). The pixel-level population density is obtained from WorldPop (\citealp{worldpop}), a large-scale producer of high-resolution age-specific population distributions for a wide array of countries.

\begin{figure}[!h]
    \centering
    \caption{{\bf Map of NMR estimates in Zambia (2009-2013) under various models.} The maps on the top row display results from unit-level Bayesian negative binomial models with BYM2 spatial effects at the second administrative level, while the maps on the bottom row display H{\'a}jek direct estimates at the first and second administrative level, respectively.  The map on the top left displays the estimates obtained from the model with spatial effects only at the second administrative level (\ref{eq:strat}), while the map on the top right displays the estimates obtained from the model with nested spatial effects (\ref{eq:stratnested}). Note that data at the second administrative level is sparse, so the bottom right map is only included to visualize the magnitude of shrinkage.}
    \includegraphics[scale=0.33]{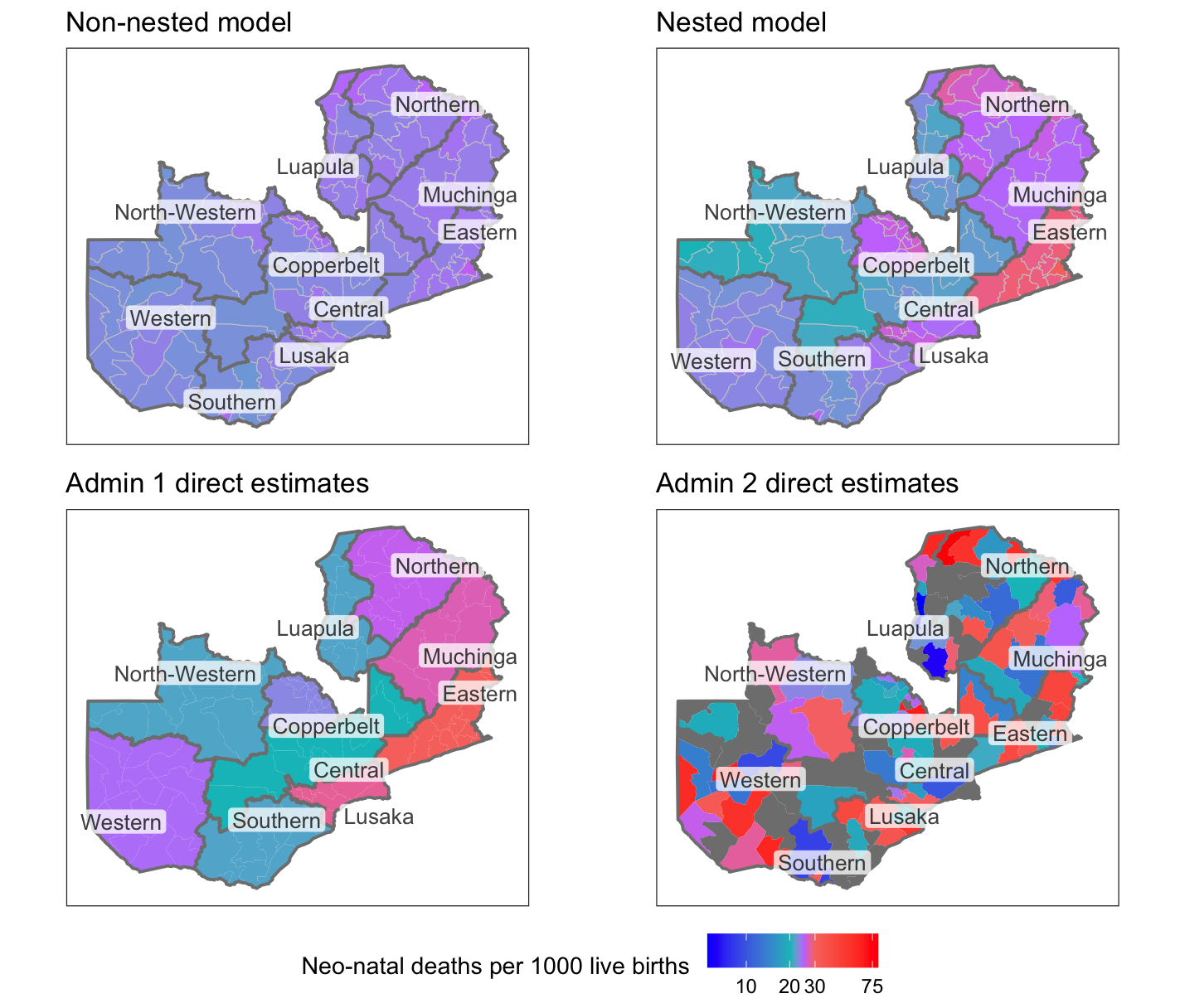}
    \label{fig:Motivating1}
\end{figure}

\begin{figure}[h!]
\caption{{\bf NMR estimates at the second administrative level in Zambia (2009-2013) under various models.} The colored circles denote the second administrative level estimates using unit-level Bayesian negative binomial models with BYM2 spatial effects at the second administrative level and the colored diamonds denote their aggregations to the first administrative level. The black diamonds denote H{\'a}jek direct estimates at the first administrative level.}
    \centering
    \hspace{-45pt}
    \includegraphics[scale=0.33]{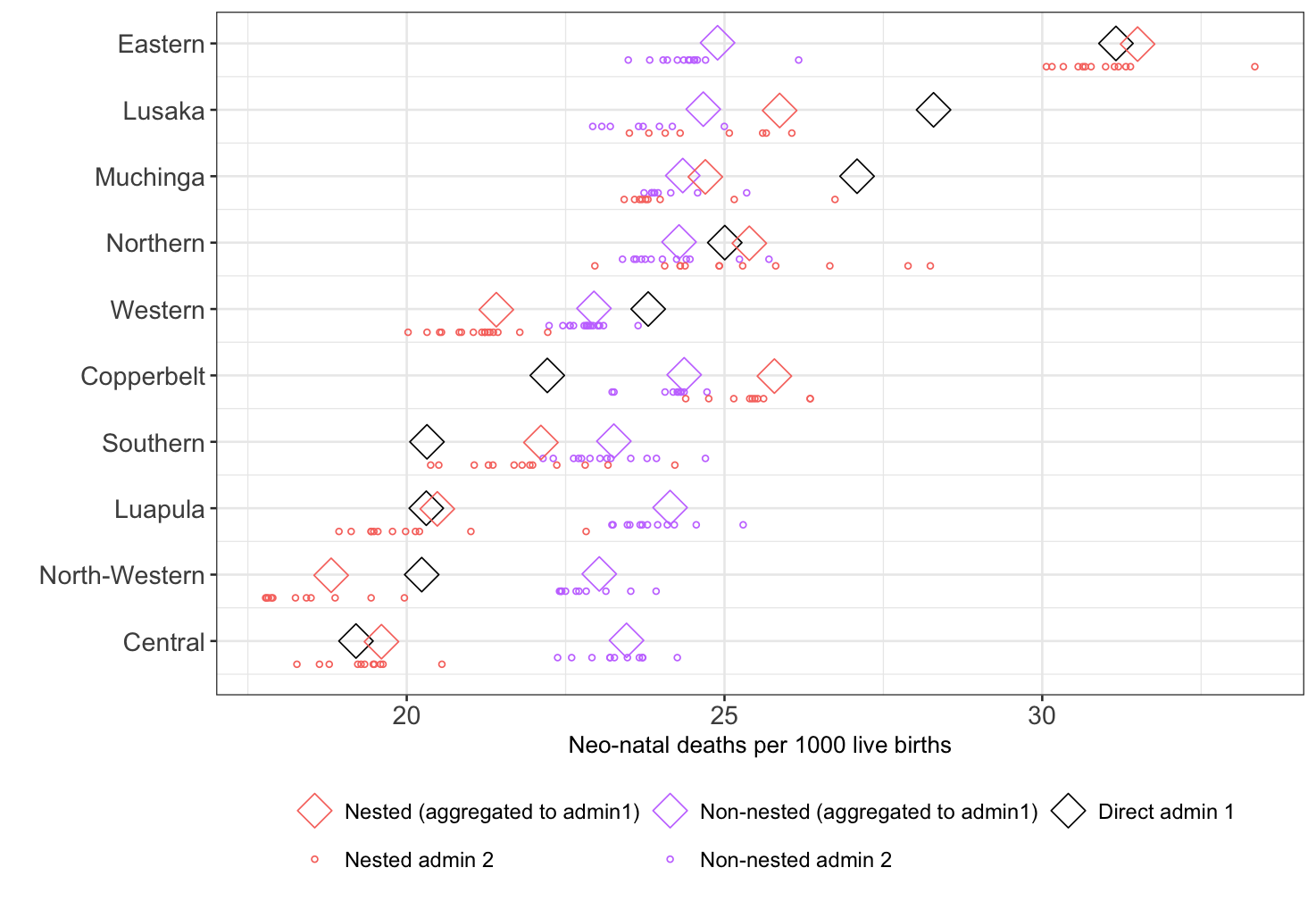}
    \label{fig:Motivating2}
\end{figure}

In figure \ref{fig:Motivating1} we map the NMR (deaths per 1000 live births) estimates under non-nested, nested, and direct methods. We observe that including a set of fixed effects at the first administrative level in the nested model significantly reduces shrinkage, by inducing shrinkage at a more local level. The nested model estimates, mapped on the top right, have a range of $18-33$ deaths per 1000 live births, while the estimates from the non-nested model, mapped on the top left, are nearly homogeneous with a range of $22-26$ deaths per 1000 live births.  In the bottom right panel of figure \ref{fig:Motivating1}, we also include a map of the second administrative level direct estimates (for areas where at least 1 neonatal death is observed) to illustrate the magnitude of shrinkage, but we caution that they are estimated from very small sample sizes and have large uncertainty (of the 87 out of 115 areas which have valid estimates, $70\%$ of areas have coefficient of variation greater than $50\%$ and $23\%$ have coefficient of variation greater than $100\%$). In \ref{CV} we provide maps of the coefficients of variation for the estimates in figure \ref{fig:Motivating1}. To further compare the model-based and direct estimates, we aggregate the second administrative level model-based estimates to the first administrative level with population weights calculated using WorldPop data (\citealp{worldpop}). In figure \ref{fig:Motivating2}, observe that the aggregation of the estimates to the first administrative level can be quite far from the consistent design-based estimate. This figure further highlights the overwhelming shrinkage of estimates under the non-nested model.

\noindent In some first administrative areas, such as the Eastern, Northern, Luapala, and Central regions, aggregated estimates from the nested model are in close agreement with their corresponding design-based estimates. However, for other areas, such as the Lusaka, Muchinga, Western, and Copperbelt regions, aggregated estimates from the nested model are quite different from their corresponding design-based estimates. While employing nested spatial effects can mitigate shrinkage in unit-level models, it is not sufficient to achieve small area estimates which agree in aggregation with higher-level design-based estimates. In fact, there is no guarantee that aggregated model-based estimates will be at all similar to their design-based counterparts. Because design-based estimators are consistent, it is desirable to consider a model which will encourage consistency with them in aggregation.

\section{\MakeUppercase{Direct-assisted Bayesian unit-level model}}

In the following section we introduce the DABUL models: an extension of the standard unit-level Bayesian model with nested random effects which incorporates direct design-based estimates to encourage consistency with design-based estimates in aggregation.

First, consider a model for a full enumeration of the population in which all births in all clusters are observed. Let $\bf{Y}$ be the vector containing the total number of neonatal deaths in all households, $Y_c$, in each cluster $c$, so that the vector has length equal to the total number of clusters in the population. Let $\bf{Y_i}$ be the sub-vector containing the elements of $\bf{Y}$ corresponding to all clusters in the first administrative area $i$, and $Y_{i+}$ be the sum of the elements in this sub-vector. Similarly, let $\bf{N}$ be the vector containing the total number of births in all households, $N_c$, in each cluster $c$, so that the sum of the elements of $\bf{N}$ is the total number of births in the population. Let $\bf{N_i}$ be the sub-vector containing the elements of $\bf{N}$ corresponding to all clusters in the first administrative area $i$, and $N_{i+}$ be the sum of the elements in this sub-vector. Also, let $\bf{r}$ be the vector containing elements $r_c$, and $\bf{r_i}$ be the sub-vector containing the elements of $\bf{r}$ corresponding to clusters in the first administrative area, $i$. Let $\odot$ denote the Hadamard product (elementwise multiplication; see \ref{notation appendix} for more details). Then an equivalent way to define a negative binomial distribution is 

\begin{equation}
P({\bf Y_i}|{\bf r_i},\lambda ) = P({\bf Y_i}, Y_{i+}|{\bf r_i},\lambda ) = P({\bf Y_i}|Y_{i+},{\bf r_i},\lambda )\times P(Y_{i+}|{\bf r_i},\lambda )
\end{equation}

\noindent for clusters in first administrative area, $i$, where

\begin{equation}
\hspace{7mm} {\bf Y_i}|Y_{i+},{\bf r_i},\lambda \sim \mbox{DCM}\left(Y_{i+}, {\bf N_i}\odot{\bf r_i}/\lambda\right)
\label{eq:DCM1}
\end{equation}

\noindent and 

\begin{equation}
   Y_{i+}|{\bf r_i},\lambda \sim \mbox{Negative-binomial}\left(\sum_{c\in\delta_1(i)}N_cr_c,\lambda\right).
\label{eq:Ypluslikelihood}
\end{equation}

Here DCM abbreviates the Dirichlet compound multinomial distribution \mbox{(\citealp{dcm})}, also referred to as the multivariate P\'olya distribution , which is defined by the probability mass function,

$$P({\bf Y_i}|Y_{i+},{\bf r_i},\lambda ) = \frac{\Gamma(Y_{i+}+1)\Gamma(\frac{1}{\lambda}\sum_{c\in\delta_1(i)} N_cr_c)}{\Gamma(Y_{i+}+\frac{1}{\lambda}\sum_{c\in\delta_1(i)} N_cr_c)}\prod_{c\in \delta_1(i)}\frac{\Gamma(Y_c + N_cr_c/\lambda )}{\Gamma(N_cr_c/\lambda )\Gamma(Y_c+1)}.$$

This alternative expression follows directly from the additive property of negative binomial random variables and the relationship between the negative binomial and DCM distributions. Specifically, if ${\bf x}=(x_1,...,x_n)$ are independent random variables, each following a negative binomial distribution with mean parameter $\boldsymbol{\theta} 
 = (\theta_1,...,\theta_n)$, respectively, and a common overdispersion parameter $\lambda$, then $x_+:=\sum_{i=1}^nx_i$ follows a negative binomial distribution with mean $\sum_{i=1}^n\theta_i$, and overdispersion parameter $\lambda$. It also holds that ${\bf x}|x_+,\boldsymbol{\theta},\lambda$ follows a DCM distribution with parameter vector $\boldsymbol{\theta}/\lambda$.

 When all births and neonatal deaths in each cluster are observed (i.e. ${\bf Y}$ and $Y_{i+}$ is known), this alternative parameterization of the negative binomial distribution is equivalent to the parameterization presented in (\ref{eq:negbinparam}). We will now consider a survey sampling context in which 1) we use two-stage sampling, so that only some clusters are sampled and only some of the births and corresponding neonatal deaths are observed in each of those sampled clusters, i.e., ${\bf Y}$ is not known, and 2) each $Y_{i+}$ can be estimated through a design-based estimate.

\subsection{Accounting for survey sampling}

 We first assume $Y_{i+}$ is known for all $i$, so that only point 1) needs to be considered. Suppose the birth and death observations are collected using the two-stage cluster sampling design described in section 2. In this situation, we must incorporate the fact that not all births, and corresponding deaths, are observed in each sampled cluster. For each cluster $c$, let $\gamma_c$ be an indicator variable denoting whether the cluster was sampled and let $n_c$ and $Z_c$ be the number of births and neonatal deaths across all of the sampled households in that cluster. Also, let the vectors $\boldsymbol{\gamma}$, $\boldsymbol{\gamma_i}$, ${\bf Z}$ and ${\bf Z_i}$ be defined analogously to ${\bf Y}$ and ${\bf Y_i}$. Note that if $\gamma_c=0$ for a cluster $c$, this implies $n_c=Z_c=0$ because that cluster was not sampled. Observe then, for each cluster $c$,

\vspace{-8mm}

\begin{equation}
Z_c|Y_c,\gamma_c=1\sim\mbox{Hypergeometric}\left(N_c,Y_c,n_c\right), \hspace{10mm} P(Z_c|\gamma_c=0) = I(Z_c=0).
\label{eq:hypergeom}
\end{equation}

\subsection{Incorporating higher level design-based estimates}

As alluded to in the previous section, in this survey sampling context each $Y_{i+}$ is not known, and must be estimated. We have established that the DABUL models pertain to the situation in which there is sufficient data to use design-based estimates at the first administrative level, though not at the second administrative level. We will denote the design-based prevalence estimate for a first administrative area $i$ as $r^{D_1}_i$. The only difference between the two DABUL models is how the design-based estimates are incorporated into the likelihood. For the hard DABUL model, we can simply set $Y_{i+} = r^{D_1}_i\times N_{i+}$, to apply a hard benchmarking constraint. For the soft DABUL model, we account for the uncertainty of the design-based estimate by using the property that design-based estimators are asymptotically normal (\citealp{asymptotics}, Section 3). On the logit scale, we can specify

\begin{equation}
    \mbox{logit}(r^{D_1}_i)|Y_{i+}\sim\mathcal{N}\left(\mbox{logit}\left(Y_{i+}/N_{i+}\right),V^{D_1}_i\right)
    \label{eq:asympdist}
\end{equation}

\noindent where $V^{D_1}_i$ is the design-based variance of logit$(r^{D_1}_i)$, which accounts for the uncertainty induced by the sampling design. This variance can be estimated using the \texttt{svydesign} function in the \texttt{survey} package (\citealp{survey}), to estimate the prevalence, which is then transformed to the logit scale using the delta method. If this method does not produce stable variance estimates, a variance smoothing method may be employed (e.g., using generalized variance functions described in chapter 7 of \cite{Wolter2007}), but we will not consider this option here.

Then, the joint distribution of $({\bf Z},{\bf r^{D_1}})$ conditional on $({\bf Y},{\bf Y_+},\boldsymbol{\gamma})$ under the hard DABUL model can be expressed as,

\begin{eqnarray}
    P({\bf Z},{\bf r^{D_1}}|{\bf Y},{\bf Y_+},\boldsymbol{\gamma}) &=& P({\bf Z}|{\bf r^{D_1}},{\bf Y},\boldsymbol{\gamma})P({\bf r^{D_1}}|{\bf Y_+})\nonumber \\
    &\propto& \prod_{i=1}^{m_1}P({\bf Z_i}|r^{D_1}_i,{\bf Y_i},\boldsymbol{\gamma}_i)I(r^{D_1}_i=Y_{i+}/N_{i+}).
    \label{eq:datajointdisthard} 
\end{eqnarray}

\noindent Similarly, the joint distribution of $({\bf Z},{\bf r^{D_1}})$ conditional on $({\bf Y},{\bf Y_+},\boldsymbol{\gamma})$ under the soft DABUL model can be expressed as,

\vspace{-5mm}

\begin{eqnarray}
P({\bf Z},{\bf r^{D_1}}|{\bf Y},{\bf Y_+},\boldsymbol{\gamma}) &=& P({\bf Z}|{\bf r^{D_1}},{\bf Y},\boldsymbol{\gamma})P({\bf r^{D_1}}|{\bf Y_+})\nonumber\\
  &=& \prod_{i=1}^{m_1}P({\bf Z_i}|r^{D_1}_i,{\bf Y_i},\boldsymbol{\gamma}_i)P(r^{D_1}_i|Y_{i+}).
  \label{eq:datajointdist} 
\end{eqnarray}

\noindent By definition, $r^{D_1}_i$ is a weighted sum of ${\bf Z_i}$, so it follows that $P({\bf Z_i}|r^{D_1}_i,{\bf Y_i},\boldsymbol{\gamma}_i)$ is a product of Hypergeometric distributions conditional on the survey-weighted sum of ${\bf Z_i}$, $g^{D_1}_i({\bf Z_i})$, being equal to $r^{D_1}_i$. More explicitly, for first administrative area $i$,

\begin{eqnarray}
    P({\bf Z_i}|r^{D_1}_i,{\bf Y_i},\boldsymbol{\gamma}_i) &=& \frac{P({\bf Z_i},g^{D_1}_i({\bf Z_i})=r^{D_1}_i|{\bf Y_i},\boldsymbol{\gamma}_i)}{P(g^{D_1}_i({\bf Z_i})=r^{D_1}_i|{\bf Y_i},\boldsymbol{\gamma}_i)}\nonumber\\
    &\propto& \frac{I(g^{D_1}_i({\bf Z_i})=r^{D_1}_i)}{P(g^{D_1}_i({\bf Z_i})=r^{D_1}_i|{\bf Y_i},\boldsymbol{\gamma}_i)}\underbrace{\left[\prod_{\substack{c\in\delta_1(i):\\\gamma_c=1}}P(Z_c|Y_c,\gamma_c=1)\right]}_{\text{Hypergeometric (\ref{eq:hypergeom})}}.
\end{eqnarray}

\noindent Note that $g^{D_1}_i({\bf Z_i})=r^{D_1}_i|{\bf Y_i},\boldsymbol{\gamma}_i$ follows the distribution of a weighted sum of Hypergeometric random variables.

\subsection{Combining likelihood components}

The joint distribution of $({\bf Z}, {\bf r^{D_1}}, {\bf Y_+}, {\bf Y})$ conditional on $(\boldsymbol{\gamma},{\bf r}, \lambda )$ can be expressed as,

\vspace{-5mm}

\begin{eqnarray}
    P({\bf Z},{\bf r^{D_1}}, {\bf Y_+},{\bf Y}|\boldsymbol{\gamma},{\bf r}, \lambda ) &\propto&  \prod_{i=1}^{m_1}P({\bf Z},{\bf r^{D_1}}|{\bf Y},{\bf Y_+},\boldsymbol{\gamma})P(Y_{i+},{\bf Y_i}| {\bf r_i},\lambda )\nonumber\\
    &\propto&\prod_{i=1}^{m_1}\underbrace{P({\bf Z},{\bf r^{D_1}}|{\bf Y},{\bf Y_+},\boldsymbol{\gamma})}_\text{(\ref{eq:datajointdisthard}) or (\ref{eq:datajointdist})}
    \underbrace{P({\bf Y_i}|Y_{i+},{\bf r_i},\lambda )}_\text{DCM (\ref{eq:DCM1})}\underbrace{P(Y_{i+}|{\bf r_i},\lambda )}_\text{Neg-bin (\ref{eq:Ypluslikelihood})}
    \label{eq:jointdist}
\end{eqnarray}

\noindent where ${\bf Y^{(s)}}$ is the sub-vector of ${\bf Y}$ containing the elements corresponding to sampled clusters and $m_1$ is the number of first administrative areas. 

Observe that in this joint distribution, the latent variables corresponding to the unobserved clusters (i.e., ${\bf Y}\setminus {\bf Y^{(s)}}$) are only included in the DCM distribution. Estimation of these latent variables will cause an identifiability issue because there is no data for these clusters, so the total neonatal death counts from all of the unobserved clusters in a first administrative area can be collapsed into one group. Because we can estimate ${\bf Y}^{(s)}$ and ${\bf Y_+}$, this collapsing ensures identifiability.  Hence, we replace (\ref{eq:DCM1}) with

\begin{equation}
({\bf Y_i^{(s)}},Y_{i+} - Y_{i+}^{(s)})|Y_{i+}, {\bf r_i},\lambda \sim \mbox{DCM}\left(Y_{i+},\left(({\bf N_i^{(s)}\odot r_i^{(s)}})/\lambda,\left(\sum_{\substack{c\in\delta_1(i) \\ \gamma_c=0}}N_cr_c\right)/\lambda\right)\right)
\end{equation}

\noindent where $Y_{i+}^{(s)}$ is the sum of the elements in ${\bf Y_i^{(s)}}$, and ${\bf N}_i^{(s)}$ and ${\bf r}_i^{(s)}$ are the sub-vectors of ${\bf N_i}$ and $\bf{r}_i$ containing elements corresponding to the observed clusters in first administrative area $i$. Note that this distribution has dimension equal to the total number of observed clusters plus one, as does the corresponding probability vector. By making this adjustment we do not lose any information and reduce the dimension of the latent variable vector from the total number of clusters, to the total number of \textit{observed} clusters, which is of significantly smaller magnitude. This reduction increases numerical stability and computational efficiency.

\begin{figure}[h!]
    \caption{{\bf Comparison of DABUL models with the standard unit-level Bayesian model.} The cluster-length vector ${\bf Y}$, indexed by $c$, denotes the total number of neonatal deaths in each cluster. The vector ${\bf Y^{(s)}}$ is the sub-vector of ${\bf Y}$ containing the elements corresponding to sampled clusters and ${\bf Z}$ is the vector of observed deaths in each sampled cluster, indexed by $c$.}
    \centering
    \includegraphics[scale=0.53]{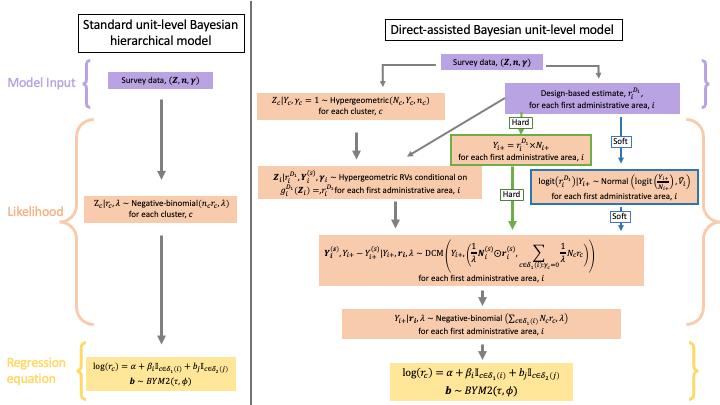}
    \label{fig:FlowChart}
\end{figure}

This completes the derivation of the posterior distribution under the DABUL models, which can be expressed as follows,

\begin{equation}
    \pi({\bf r}, \lambda, \boldsymbol{\kappa}, {\bf Y^{(s)}},{\bf Y_+}|{\bf Z},{\bf r^{D_1}},\boldsymbol{\gamma}) 
    \propto \underbrace{P({\bf Z},{\bf r^{D_1}}, {\bf Y_+},{\bf Y}|\boldsymbol{\gamma},{\bf r}, \lambda )}_\text{(\ref{eq:jointdist})}P({\bf r}|\boldsymbol{\kappa})\pi(\lambda,\boldsymbol{\kappa})
\label{eq:DABUL}
\end{equation}

\noindent where $\boldsymbol{\kappa}$ is the set of regression parameters and hyperparameters. Figure \ref{fig:FlowChart} depicts the components of this model in a visual form and compares it to the standard unit-level Bayesian model with nested spatial effects. Observe the only difference between the hard and soft DABUL models is how $\bf{{r}^{D_1}}$ is used to estimate $\bf{Y_+}$.

\subsection{Implementation}

Estimation of the posterior distribution (\ref{eq:DABUL}) is not straightforward because most Bayesian inference software, including \texttt{Stan} (\citealp{standev}) and the \texttt{INLA} package in \texttt{R} (\citealp{inla}), are not built to accommodate discrete latent variables that cannot be marginalized out. As a result, we implement a NUTS-within-Gibbs sampler, described in Algorithms \ref{alg:main} and \ref{alg:hard}, which uses a No-U-Turn Sampler (NUTS) to update all continuous parameters and inverse transform sampling to update the discrete parameters. Introduced by \cite{NUTS}, NUTS is an efficient extension of Hamiltonian Monte Carlo and is the computational framework for \texttt{Stan}.  We implemented the NUTS step of our algorithm in R according to Algorithm 3 in \cite{NUTS}, with assistance from the code provided by \cite{NUTSexample}. We modified the algorithm to include a scaling matrix, denoted $\boldsymbol{\Sigma}$, in the proposal density to better account for correlation between parameters, as is done in \texttt{Stan}. The gradient of the log-likelihood with respect to all continuous parameters was derived analytically to avoid costly numerical approximations at each step of the NUTS sampler. The tuning parameter, $\epsilon$, is chosen to optimize the step size of the random walk.

All components of the posterior distribution (\ref{eq:DABUL}) have a closed form except for $P(g^{D_1}_i({\bf Z_i})=r^{D_1}_i|{\bf Y_i},\boldsymbol{\gamma}_i)$, the probability mass function of a weighted sum of independent Hypergeometric random variables. Although $g^{D_1}_i({\bf Z_i})|{\bf Y_i},\boldsymbol{\gamma}_i$ follows a discrete distribution, the domain contains all the possible weighted sums of ${\bf Z}_i$, which is very dense because each $Z_c$ has a different corresponding weight. Therefore, in practice, we can treat $P(g^{D_1}_i({\bf Z_i})=r^{D_1}_i|{\bf Y_i},\boldsymbol{\gamma}_i)$ as a probability density function (with continuous domain). At each iteration of the algorithm this component must be approximated for the inverse transform sampling of ${\bf Y}^{(s)}$ step. We approximate this term by drawing 2000 multivariate samples from the independent Hypergeometric distributions implied by the proposed and current states and taking the weighted sum of each sample vector, resulting in 2000 draws from the distribution of $g^{D_1}_i({\bf Z_i})|{\bf Y_i},\boldsymbol{\gamma}_i$. Gaussian kernel density estimation is then applied to this empirical distribution, using the \texttt{density} function in \texttt{R}, to approximate $P(g^{D_1}_i({\bf Z_i})=r^{D_1}_i|{\bf Y_i},\boldsymbol{\gamma}_i)$. The bandwidth is chosen using Silverman's rule-of-thumb method (\citealp{silvermankde}).

\begin{algorithm}

\caption{NUTS-within-Gibbs sampler for soft DABUL model}
\textbf{Input:} $({\bf Z},\boldsymbol{\gamma},{\bf n},{\bf N})$; $({\bf r^{D_1}}, {\bf V^{D_1}})$; $\epsilon$, $\boldsymbol{\Sigma}$, $M$, $\ell_{REG}:=\log\pi(\alpha, \boldsymbol{\beta}, {\bf b}, \sigma^2, \phi, \lambda|{\bf Y^{(s)}},{\bf Y_+},{\bf Z},\boldsymbol{\gamma})$ 

\begin{algorithmic}
\For{$m$ in 1:$M$}

\State $(\alpha, \boldsymbol{\beta}, {\bf b}, \sigma^2, \phi, \lambda )^{(m)} \longleftarrow NUTSOneStep((\alpha, \boldsymbol{\beta}, {\bf b}, \sigma^2, \phi, \lambda )^{(m-1)},\ell_{REG},({\bf Y^{(s)}},{\bf Y_+})^{(m-1)},\epsilon,\boldsymbol{\Sigma})$

\ForEach{cluster, $c$}
\State $r_c^{(m)} \longleftarrow exp(\alpha + \beta_i\mathds{1}_{c \in \delta_1(i)} + b_j\mathds{1}_{c \in \delta_2(j)})$
\EndFor

\ForEach{first administrative area, $i$}
    \State Draw $Y_{i+}^{(m)}\sim P(Y_{i+}| {\bf Y^{(s)}}^{(m-1)},\boldsymbol{\gamma},{\bf r}^{(m)},\lambda^{(m)},r^{D_1}_i, V^{D_1}_i)$ using inverse transform sampling
\EndFor

\ForEach {observed cluster, $c$}
    \State Draw $Y_c^{(m)}\sim P(Y_c| {\bf Y_{1:(c-1)}^{(s)}}^{(m)}, {\bf Y^{(s)}_{(c+1):n_{clusters}}}^{(m-1)},Y_{i+}^{(m)},Z_c,\gamma_c,r_c^{(m)},\lambda^{(m)})$ using inverse transform sampling
\EndFor

\EndFor

\end{algorithmic}

\textbf{Output:} $\alpha, \boldsymbol{\beta}, {\bf b}, \lambda,\sigma^2, \phi, {\bf Y^{(s)}}, {\bf Y_+}$

\label{alg:main}

\end{algorithm}

\begin{algorithm}[h!]
\caption{NUTS-within-Gibbs Sampler for hard DABUL model}
\textbf{Input:} $({\bf Z},\boldsymbol{\gamma},{\bf n},{\bf N})$; ${\bf r^{D_1}}$; $\epsilon$, $\boldsymbol{\Sigma}$, $M$, $\ell_{REG}:=\log\pi(\alpha, \boldsymbol{\beta}, {\bf b}, \sigma^2, \phi, \lambda|{\bf Y^{(s)}},{\bf Y_+},{\bf Z},\boldsymbol{\gamma})$

\begin{algorithmic}
\State $Y_{i+} = r^{D_1}_iN_{i+}$ for each first administrative area $i$
\For{$m$ in 1:$M$}

\State $(\alpha, \boldsymbol{\beta}, {\bf b}, \sigma^2, \phi, \lambda )^{(m)} \longleftarrow NUTSOneStep((\alpha, \boldsymbol{\beta}, {\bf b}, \sigma^2, \phi, \lambda )^{(m-1)},\ell_{REG},{\bf Y^{(s)}}^{(m-1)},{\bf Y_+},\epsilon,\boldsymbol{\Sigma})$

\ForEach {sampled cluster, $c$}
\State $r_c^{(m)} \longleftarrow exp(\alpha + \beta_i\mathds{1}_{c \in \delta_1(i)} + b_j\mathds{1}_{c \in \delta_2(j)})$
    \State Draw $Y_c^{(m)}\sim P(Y_c| {\bf Y^{(s)}_{1:(c-1)}}^{(m)}, {\bf Y^{(s)}_{(c+1):n_{clusters}}}^{(m-1)},Y_{i+},Z_c,\gamma_c,r_c^{(m)},\lambda^{(m)})$ using inverse transform sampling
\EndFor

\EndFor

\end{algorithmic}
\textbf{Output:} $\alpha, \boldsymbol{\beta}, {\bf b}, \lambda,\sigma^2, \phi, {\bf Y^{(s)}}$

\label{alg:hard}
\end{algorithm}

\section{\MakeUppercase{Simulation Study}}

\begin{wrapfigure}{r}{0.5\textwidth}
\vspace{-30pt}
\caption{{\bf Second administrative area neighborhood structure for simulation study.}}
\includegraphics[width=0.5\textwidth]{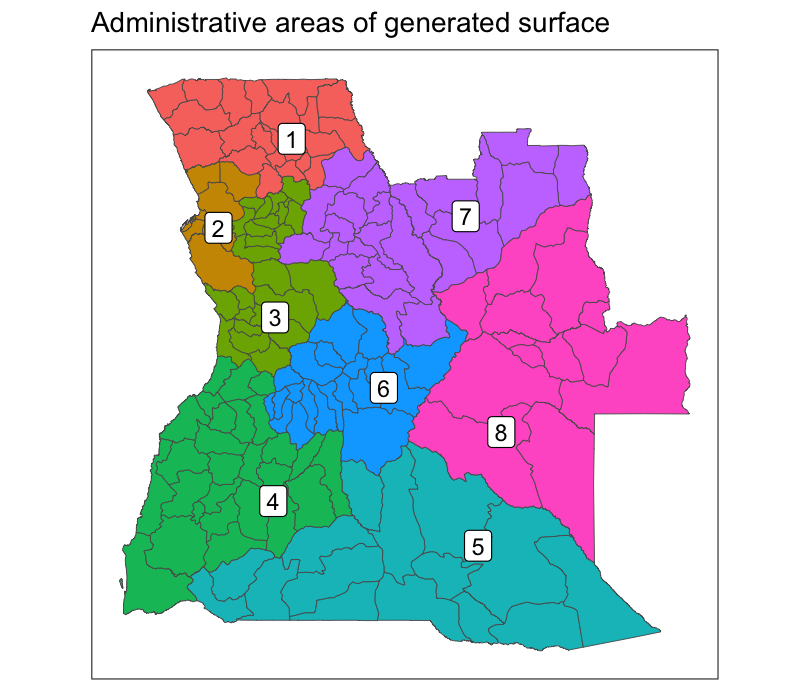}
\vspace{-20pt}
\label{fig:SimMap}
\end{wrapfigure}

In this study we will compare the performance of the proposed DABUL models to an analogous, standard unit-level Bayesian model with nested spatial effects.  We start by defining a neighborhood structure with 8 first administrative areas and 159 second administrative areas. There is a minimum number of 13 second administrative areas in each first administrative area and a maximum of 27. A map of these areas is provided in figure \ref{fig:SimMap}. This neighborhood structure and map is modified from that of Angola. The total number of urban and rural clusters in each first administrative area is drawn from $\mathcal{N}(700,100^2)$ and $\mathcal{N}(800,100^2)$ distributions, respectively, and the total number of births in each urban cluster and each rural cluster is drawn from $\mathcal{N}(100,10^2)$ and $\mathcal{N}(125,10^2)$ distributions, respectively. All draws are rounded to the nearest natural number. These values are calibrated to be comparable to the sampling frames in LMICs used by DHS and MICS. Within each first administrative area, the clusters are distributed evenly across second administrative areas so that the number of births and clusters in a second administrative area are similar to other second administrative areas within its first administrative area.

The total number of neonatal deaths in each cluster, $Y_c$, is drawn from a negative binomial distribution with mean $N_cr_c$ and overdispersion parameter $\lambda=0.25$, where $r_c$ is defined as

\begin{equation}
    \log(r_c) =\alpha_U\mathds{1}_{c \in 
    \delta_U} + \alpha_R\mathds{1}_{c \in 
    \delta_R} + \beta_i\mathds{1}_{c \in \delta_1(i)} + b_j\mathds{1}_{c \in \delta_2(j)}
    \label{eq:regeqn}
\end{equation}

\noindent where $\alpha_U = \log(0.02), \alpha_R=\log(0.025)$, and $\boldsymbol{\beta} =(-0.2,-0.1,-0.05,-0.025,0.025,0.05,0.1,0.5)$. Separate urban and rural intercepts are specified to replicate real-world conditions in which prevalence are often different between urban and rural areas. This corresponds to a neonatal mortality rate between 16 and 41 per 1000 live births, which figure \ref{fig:Motivating1} shows is a reasonable range. Additionally neonatal mortality rate estimates in many Sub-Saharan African countries, including Zambia, which are available on the UN-IGME Child Mortality Estimation dashboard (\texttt{childmortality.org}), confirm that this is a reasonable choice. The choice of overdispersion parameter $d=0.25$ is also reasonable compared to the overdispersion parameter estimate in the motivating Zambia example, $0.19$ ($90\%$ credible interval: $[0.08, 0.32]$).  The random spatial effects ${\bf b}$ are drawn from a BYM2 model with parameters $\sigma^2$ and $\phi$, which is equivalent to a mean-zero multivariate normal distribution with covariance matrix, $\sigma^2((1-\phi){\bf I} + \phi{\bf Q^-_*})$, where ${\bf Q^-_*}$ is the unique Moore-Penrose generalized inverse of the scaled structure matrix, ${Q_*}$, as described in \cite{BYM2}. The values of $\sigma^2$ and $\phi$ vary over three different hyperparameter settings: in the first, $\sigma^2=0.15^2$ and $\phi=0.25$; in the second, $\sigma^2=0.05^2$ and $\phi=0.25$; and in the third $\sigma^2=0.05^2$ and $\phi=0.7$. The choices of $\sigma$ are reasonable compared to the $\sigma$ parameter estimate in the motivating Zambia example, $0.12$ ($90\%$ credible interval: $[0.03, 0.31]$). For the choice of $\phi$, anywhere in the range of $(0,1)$ is reasonable, so we pick two distinct values in that range. Through these settings, we can observe whether the relative performance of the DABUL model is affected by spatial precision or dependence.

\singlespacing

\doublespacing

For each simulation setting, 500 datasets are independently sampled from a single generated risk surface using two-stage stratified cluster sampling (the same method which is used by DHS and MICS, except that in the second stage we directly sample births instead of households). A summary of the four simulation settings is provided in table \ref{tab:tab2}. A proportion of urban and rural clusters are sampled from each first administrative area. For the majority of simulation settings, 8\% and 5\% of urban and rural clusters are sampled, respectively, but for one setting, only 5\% and 3\% are sampled so that the effect of sample size can be observed. Note that urban clusters are oversampled as this is often the case for DHS and MICS. Similarly, note that clusters are sampled at the first administrative level to mimic DHS and MICS, which are powered at the first administrative level. Due to this sampling design, there may be second administrative areas which have very small samples, or no sampled clusters at all.

\begin{table}[h]
    \centering
    \caption{Summary of simulation parameters and sampled datasets}
    \begin{tabular}{l|ccc|c}
        & \multicolumn{4}{|c}{Simulation setting}\\
        \hline
         & 1 & 2 & 3 & 1a \\
         \hline\hline
         Spatial variance: $\sigma^2$ & $0.15^2$ & $0.05^2$ & $0.05^2$ & $0.15^2$ \\
         Proportion of variation explained by structured effect: $\phi$ & 0.25 & 0.25 & 0.7 & 0.25 \\
         \hline
         Percentage urban clusters sampled & 8\% & 8\% & 8\% & 5\% \\
         Percentage rural clusters sampled &  5\% & 5\% & 5\% & 3\% \\
         \hline
         Avg. \# of births per 1st admin area & 1640 & 1640 & 1640 & 1007 \\
         Avg. \# of neonatal deaths per 1st admin area & 40 & 39 & 39 & 23 \\
         \hline
         Avg. \# of births per 2nd admin area & 83 & 83 & 83 & 51 \\
         Avg. \# of neonatal deaths per 2nd admin area & 2.0 & 2.0 & 2.0 & 1.2 \\
         Pct. 2nd admin areas w/ no observed neonatal deaths & 23.2 & 22.9 & 23.0 & 39.7 \\
    \end{tabular}
    \label{tab:tab2}
\end{table}

For each sampled urban cluster, the number of births sampled is drawn from $\mathcal{N}(15,4)$ and for each sampled rural cluster, the number of births sampled is drawn from $\mathcal{N}(20,4)$. Each birth in a cluster has equal probability of being sampled. A summary of the resulting datasets (i.e., average numbers of observed births and deaths per administrative area) is displayed in table \ref{tab:tab2}. For each of these 2000 sample datasets (4 settings $\times$ 500 samples), we obtain the following:

\begin{enumerate}
    \item H{\'a}jek direct estimates and standard errors at the first administrative level using the \texttt{SUMMER} (\citealp{SUMMER}) and \texttt{survey} (\citealp{survey}) packages in R.
    \item Second administrative level prevalence estimates from a standard unit-level Bayesian model using \texttt{Stan} (\citealp{standev}), by fitting the nested BYM2 regression model specified in (\ref{eq:regeqn}) under the assumption that each $Y_c$ follows a negative binomial distribution with mean $N_cr_c$ and overdispersion parameter $\lambda$.
    \item Second administrative level prevalence estimates from the soft DABUL model with the regression equation specified in (\ref{eq:regeqn}) using Algorithm \ref{alg:main} for 1000 iterations, after 1000 iterations of burn-in, for 4 chains.
    \item Second administrative level prevalence estimates from the hard DABUL model with the regression equation specified in (\ref{eq:regeqn}) using Algorithm \ref{alg:hard} for 1000 iterations, after 1000 iterations of burn-in, for 4 chains.
\end{enumerate}

As in the motivating example, we aggregate the urban and rural estimates within each second administrative area using urban/rural population fractions. The run time of each DABUL model was approximately 130 minutes, while the run time of each standard unit-level model in \texttt{Stan} was approximately 2 minutes.

We compare the performance of the second administrative level prevalence estimates using four metrics. First, we quantify the discrepancy, or absolute difference, between the first administrative level aggregated estimates and the direct estimates, averaged across all simulations. For first administrative area $i$, the average discrepancy is calculated by $$\frac{1}{500}\sum_{k=1}^{500}|\sum_{j\in\zeta (i)}w_j\hat r_j^{(k)}-r^{D_1(k)}_i|$$ where $\hat r_j^{(k)}$ is the second administrative level prevalence estimate for the $k^{th}$ sample dataset, $w_j$ is the population weight for second administrative area $j$, and $\zeta(i)$ is the set of second administrative areas in first administrative area $i$.  Second, we evaluate the absolute error of the second administrative level estimates, averaged across all simulations, calculated by $\frac{1}{500}\sum_{k=1}^{500}|\hat r_j^{(k)}-r_j|$, where $r_j$ is the population prevalence for second administrative area $j$. Then, we compare the coverage and average width of the second administrative level 90\% credible intervals, calculated by $$\frac{1}{500}\sum_{k=1}^{500}I(r_j\in [Q_j^{(k)}(5),Q_j^{(k)}(95)])$$ and $$\frac{1}{500}\sum_{k=1}^{500}(Q_j^{(k)}(95)-Q_j^{(k)}(5)),$$ respectively, where $Q_j^{(k)}(q)$ is the $q^{th}$ quantile of the posterior distribution for second administrative area $j$ and sample dataset $k$. Additionally, coefficients of variation of the second administrative level estimates, calculated by $\frac{1}{500}\sum_{k=1}^{500}100\times \frac{SD_j^{(k)}}{\hat r_j^{(k)}}$, are presented in \ref{Appendix A}, where $SD_j^{(k)}$ is the standard deviation of the posterior distribution for second administrative area $j$ and sampled dataset $k$. 

\begin{figure}[h]
\caption{{\bf Discrepancy between aggregated model-based second administrative level estimates and design-based first administrative level estimates, across 500 simulations for each of 4 settings.}}
    \centering
    \includegraphics[scale=0.25]{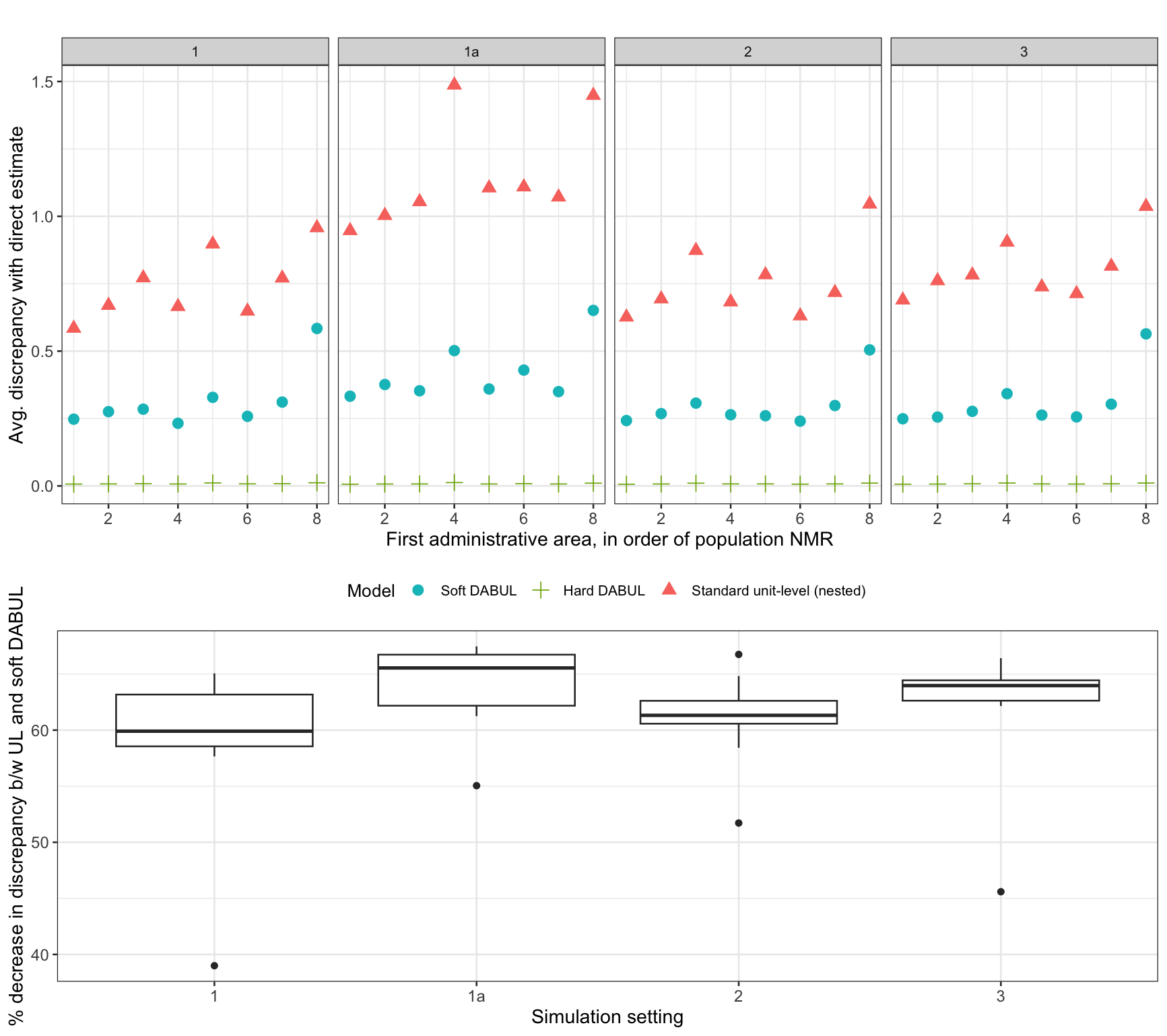}
    \label{fig:SimDisc1}
\end{figure}

Figure \ref{fig:SimDisc1} displays the discrepancies between aggregated second administrative level model-based estimates and first administrative level direct estimates. The top panel depicts the average discrepancies across simulations for each first administrative area, model, and simulation setting. From this figure we observe that discrepancies are uniformly lower among the aggregated soft DABUL estimates, compared to the aggregated standard unit-level model estimates. As expected, the discrepancies among the aggregated hard DABUL estimates are consistently near zero, as a result of enforcing a hard benchmarking constraint.  Both of these observations are consistent across all four simulation settings.  The bottom panel depicts the distribution of the average percent decrease in discrepancy of the soft DABUL estimates for each first administrative area, relative to the standard unit-level estimates, across all simulations, for each simulation setting. 

We observe that the aggregated soft DABUL estimates have $58.4$-$63.9\%$ lower discrepancy with the first administrative level direct estimates, on average, compared to aggregated estimates from the standard unit-level model. Percentage decrease in discrepancy is larger when the standard unit-level model has larger discrepancy with the direct estimates (scenario 1a). Figure \ref{fig:SimDisc2} presented in \ref{Appendix A} further shows this pattern.

Figure \ref{fig:SimSE} displays the absolute error of the second administrative level estimates, i.e., the absolute value of the difference between the prevalence estimate and the true prevalence in that area. The top panel depicts the average absolute error across simulations for each second administrative area, model, and simulation setting. The bottom panel depicts the distribution of the average absolute error of each second administrative area across

\begin{figure}[!h]
    \caption{{\bf Absolute error of second administrative level estimates, across 500 simulations for each of 4 settings.}}
    \centering
    \includegraphics[scale=0.3]{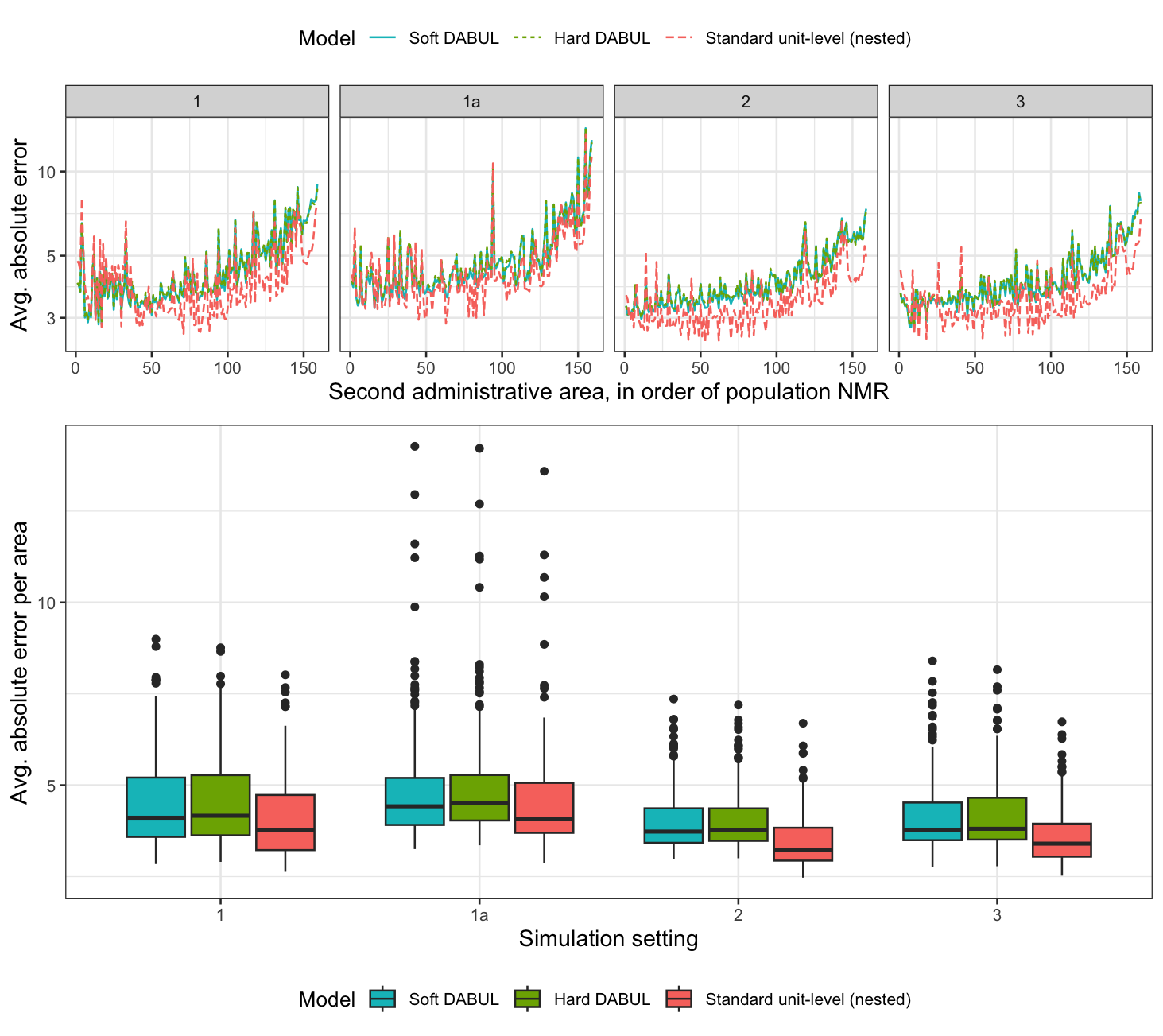}
    \label{fig:SimSE}
\end{figure}

 \noindent all simulations, for each model and simulation setting. This figure suggests that absolute error is comparable across all models and simulation settings, with a slightly lower average absolute error for standard unit-level model estimates.

\begin{figure}[!h]
\caption{{\bf Coverage of 90\% credible intervals for second administrative level estimates, across 500 simulations for each of 4 settings.}}
    \centering
    \includegraphics[scale=0.35]{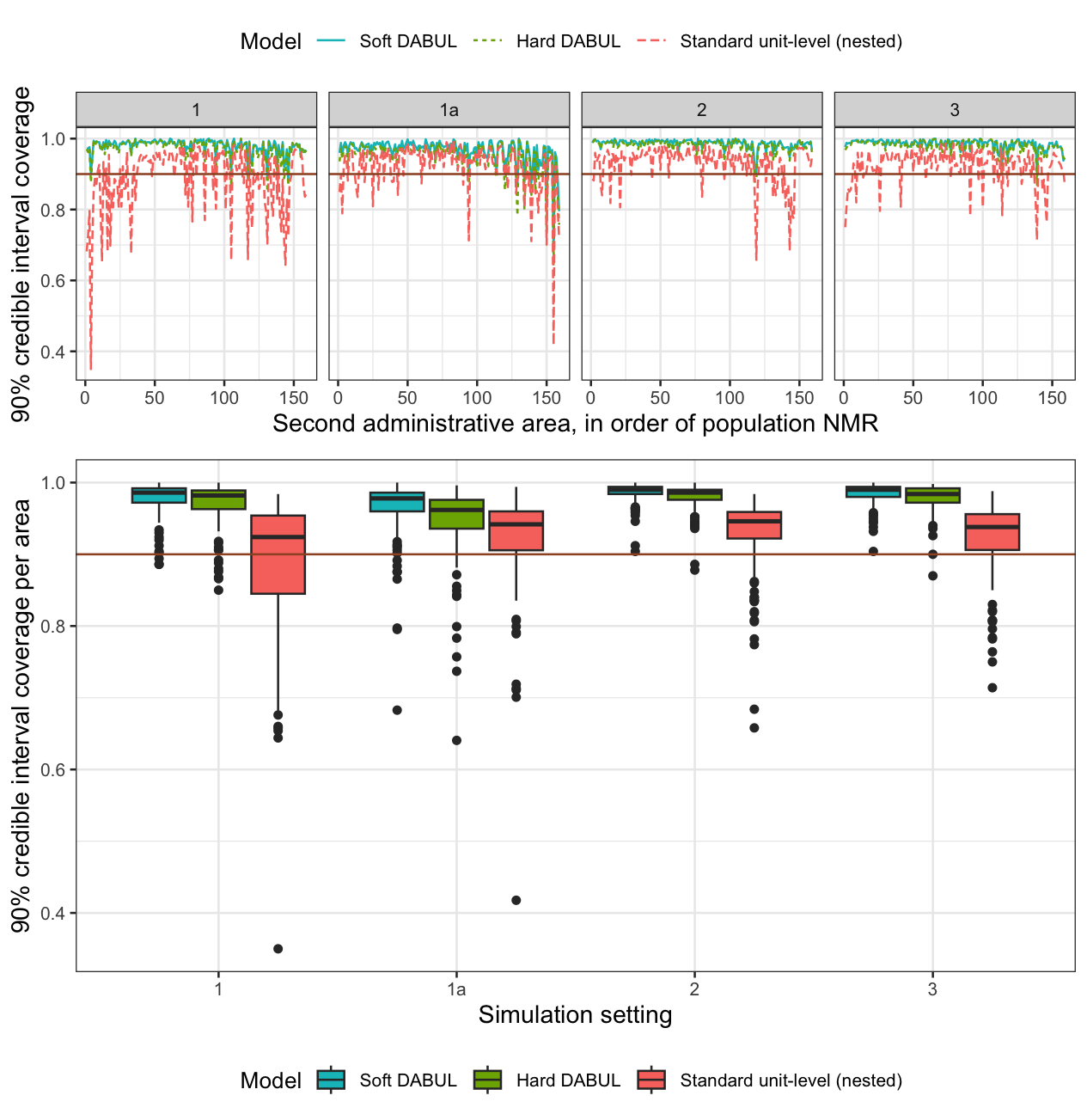}
    \label{fig:SimCoverage}
\end{figure}

\begin{figure}[h!]
\centering
    \caption{{\bf Width of 90\% credible intervals (in deaths per 1000 births) for second administrative level estimates, across
500 simulations for each of 4 settings.}}
    \includegraphics[scale=0.35]{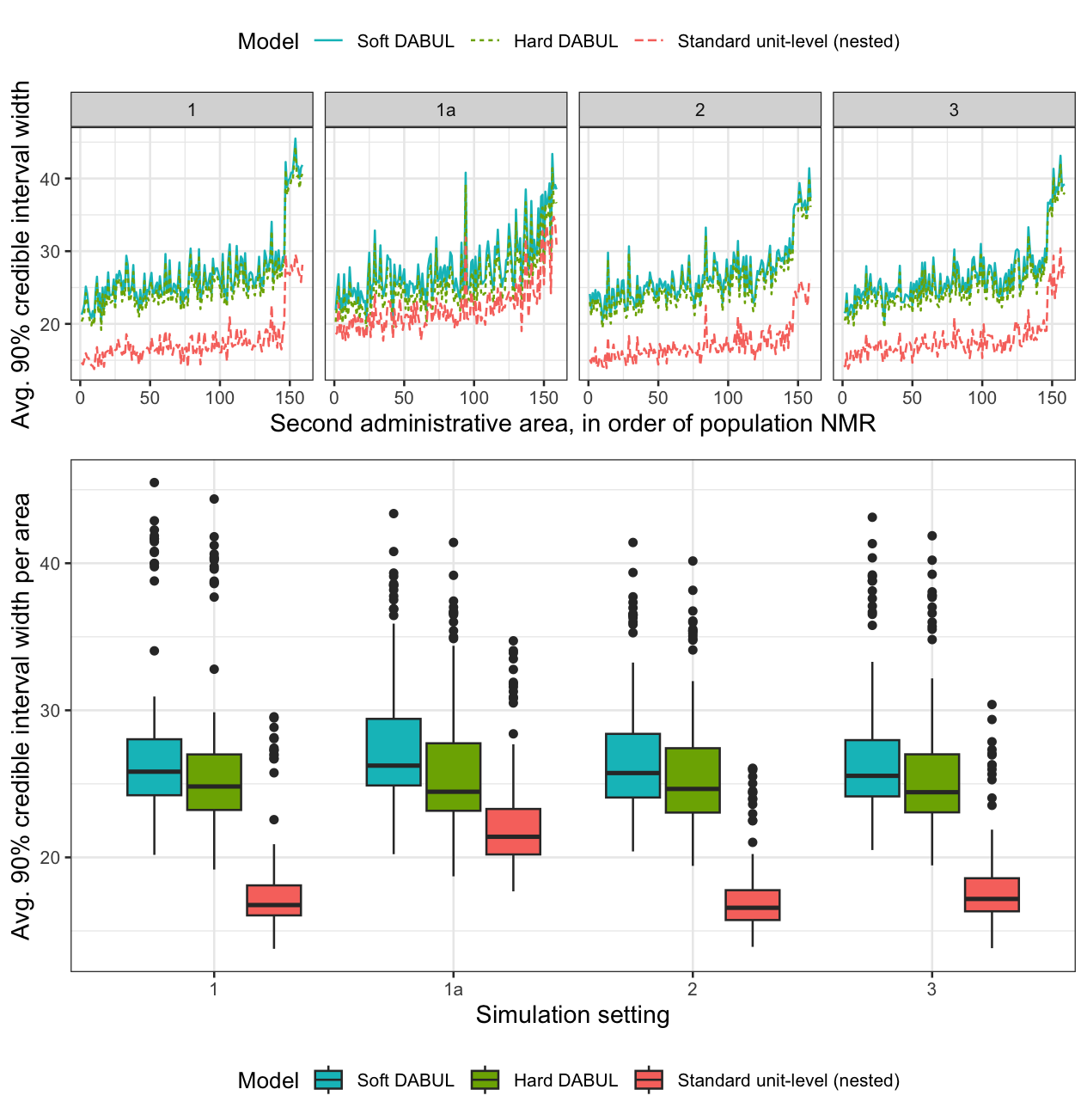}
    \label{fig:SimCIWidth}
\end{figure}

Figures \ref{fig:SimCoverage} and \ref{fig:SimCIWidth} compare the coverage and width of the 90\% credible intervals for the second administrative level estimates. Note that coverage of area-specific Bayesian credible intervals in unit-level models (and that of frequentist confidence intervals, for that matter) 

\noindent are expected to attain the desired coverage \textit{on average}, but the interval for each specific area does not necessarily have the target coverage (\citealp{yuhoff}). Moreover, in an extensive simulation study, \cite{osgood} found that coverage of area-specific credible intervals from unit-level models did not obtain the nominal level in a range of scenarios.  From figures \ref{fig:SimCoverage} and \ref{fig:SimCIWidth}, we observe that the DABUL estimates have wider credible intervals and more consistent coverage at or above the target threshold,

\noindent compared to the standard unit level model. This higher variability is due to the estimation of the additional parameters $\bf{Y}$ (and $\bf{Y_+}$ for the soft DABUL model). We observe that the coverage and width of the hard DABUL credible intervals are slightly lower than the soft DABUL credible intervals as a result of not accounting for the uncertainty of the benchmarks. While the credible intervals from the standard unit level model are narrower and have a marginal rate that is closer to the target rate, there are also significantly more areas for which the credible intervals have considerable undercoverage. From these results we observe that DABUL estimates provide more conservative credible intervals, with the benefit that a much higher proportion of areas do not have undercoverage.

\section{\MakeUppercase{Application to Zambia DHS data}}

We return to the example of NMR estimation in Zambia and compare the performance of the DABUL models to the standard unit-level model with nested spatial effects. We use the same priors and hyperpriors as in the motivating example. We use WorldPop population estimates (\citealp{worldpop}) and sampling frame information from the 2014 Zambia DHS (\citealp{dhsdata}) to obtain population counts. The run time of each of the DABUL models was approximately 90 minutes, while the run time of the standard unit-level model in \texttt{Stan} was approximately 1 minute.

\begin{figure}[h!]
    \centering
    \caption{{\bf Map of NMR estimates at the second administrative level in Zambia (2009-2013) using the DABUL models, as compared to a standard unit-level Bayesian model, both with BYM2 spatial effects at the second administrative level nested based on first administrative level.}}
    \includegraphics[scale=0.28]{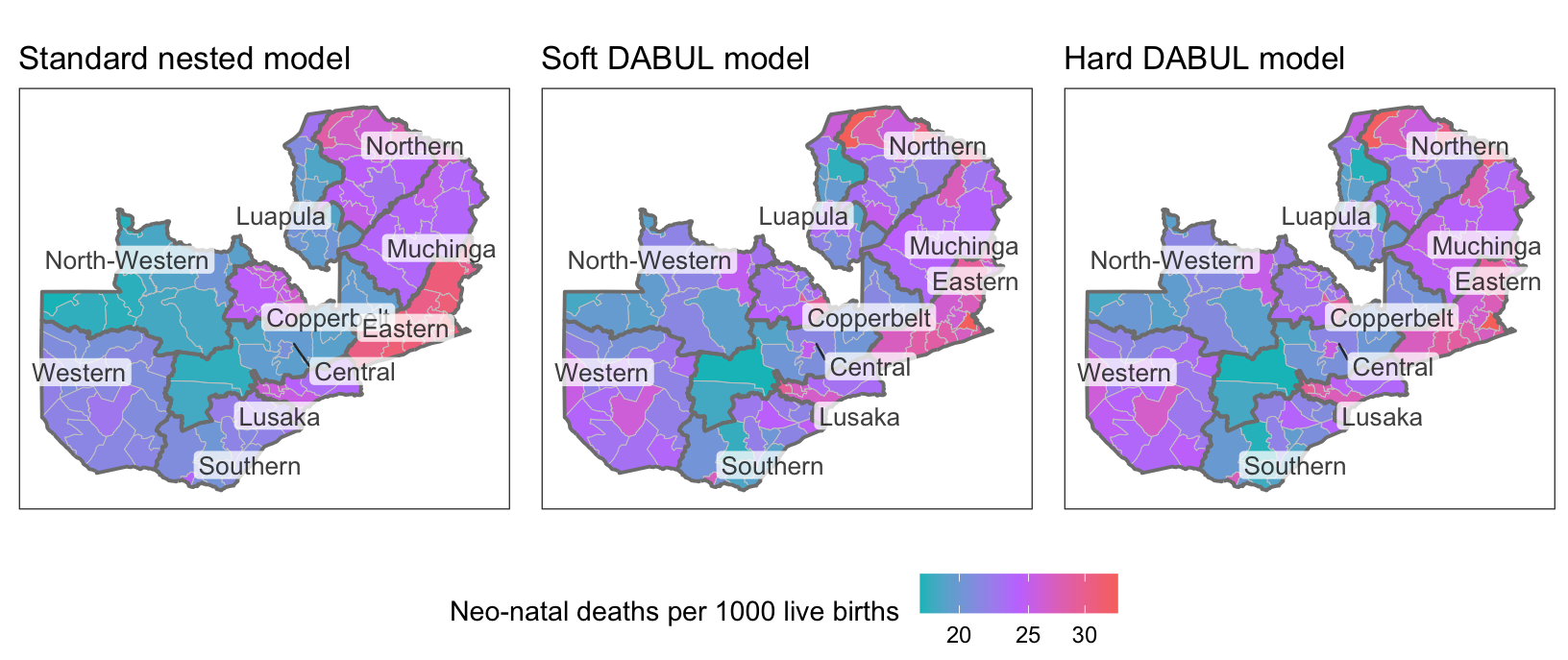}
    \label{fig:DABULmap}
\end{figure}

\begin{figure}[h!]
    \centering
    \caption{{\bf Width of 90\% credible intervals of NMR estimates (in deaths per 1000 live births) at the second administrative level in Zambia (2009-2013) using the DABUL models, as compared to a standard unit-level Bayesian model.}}
    \includegraphics[scale=0.28]{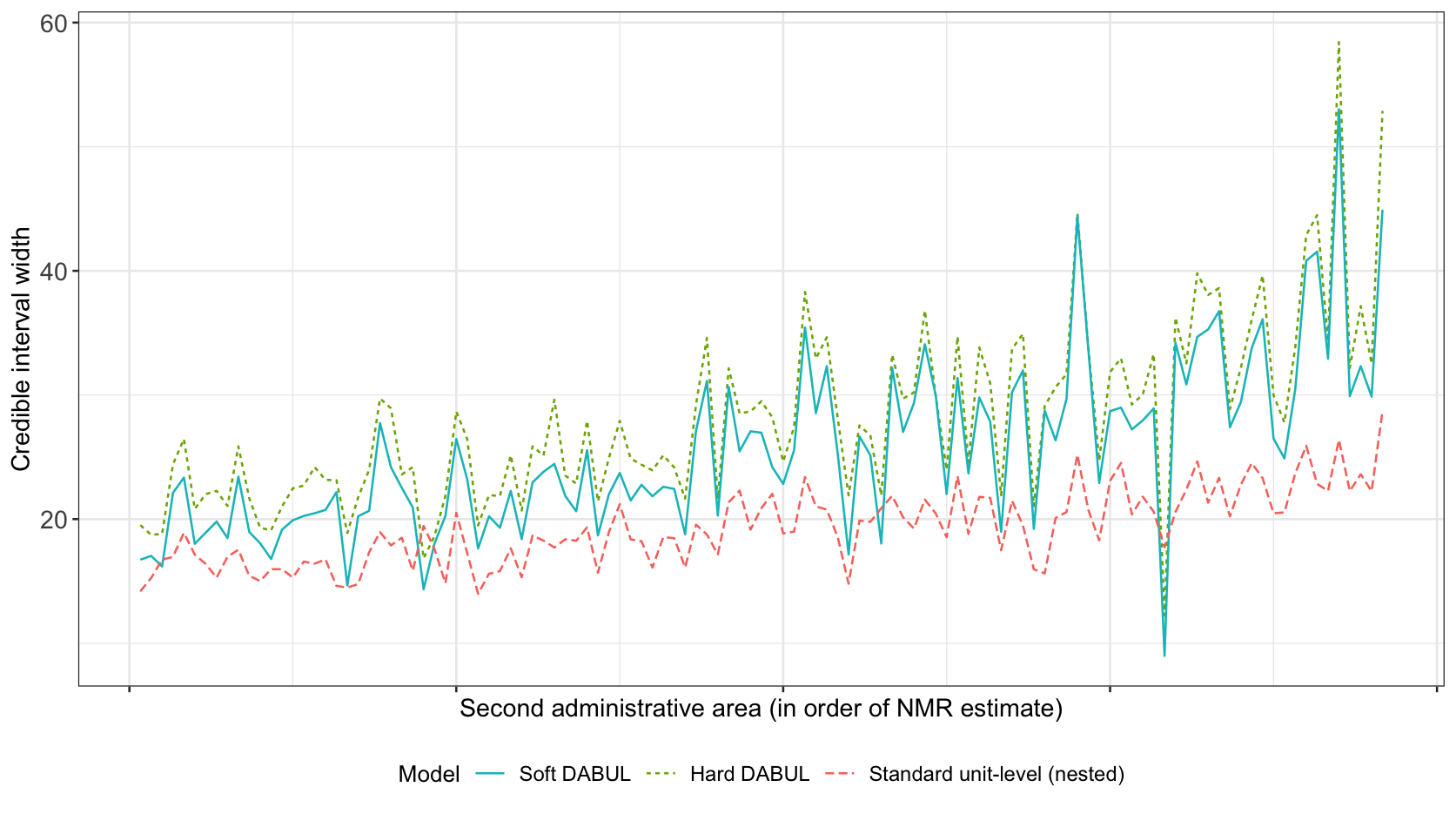}
    \label{fig:DABULciwidth}
\end{figure}

Figure \ref{fig:DABULmap} provides a map of the NMR estimates resulting from each of the three models and figure \ref{fig:DABULciwidth} compares the credible interval width for each second administrative area. A scatter plot comparing the NMR estimates from each of the DABUL models to the standard nested model is presented in \ref{Appendix B}. From these figures, we observe that the NMR point estimates are fairly similar between models, with the DABUL estimates having less shrinkage towards the first administrative level. The credible interval widths from the DABUL estimates are uniformly wider than those resulting from the standard unit-level model, which is in agreement with the simulation results.  

\begin{figure}[h!]
    \centering
    \caption{{\bf NMR estimates at the second administrative level in Zambia (2009-2013) using the soft DABUL model, as compared to a standard unit-level nested Bayesian model.} The colored circles denote the second administrative level estimates and the colored diamonds denote their aggregations to the first administrative level. The black diamonds denote H{\'a}jek direct estimates at the first administrative level.}
    \hspace{-20pt}
    \includegraphics[scale=0.3]{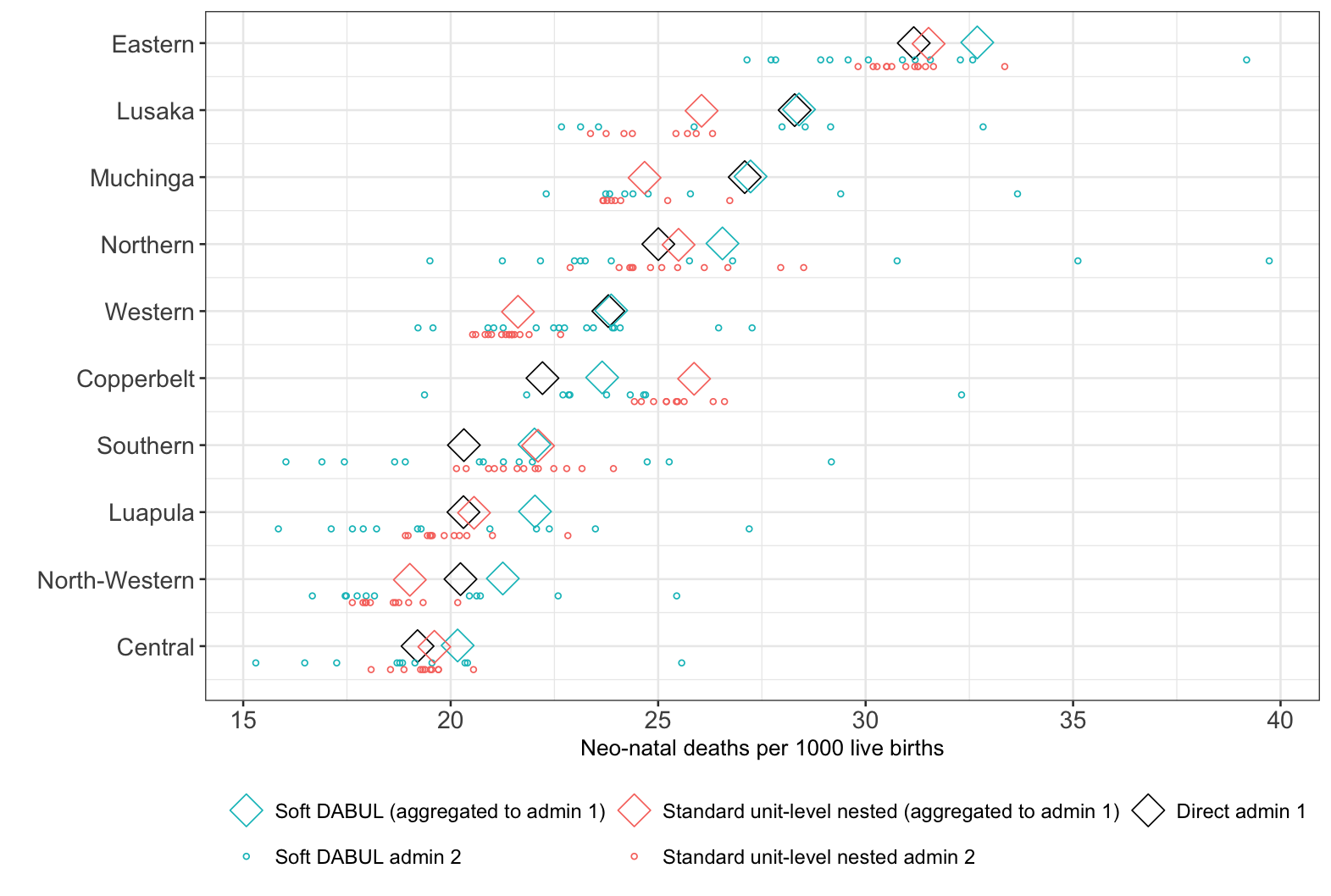}
    \label{fig:DABULagg}
\end{figure}

In figure \ref{fig:DABULagg} we evaluate whether the aggregated soft DABUL model estimates have less discrepancy with the direct estimates than the aggregated standard model estimates do. For 6 out of the 10 regions, the DABUL model estimates have similar or less discrepancy with the direct estimates, while 5 of these regions have a significant reduction in discrepancy. Two of the regions for which discrepancy increases (Luapala and Central) have substantially higher uncertainty, with coefficients of variation of $30\%$ and $31\%$, respectively, compared to the other first administrative area level direct estimates, which have coefficients of variation between $18\%$ and $23\%$. This is an expected consequence of taking the uncertainty of the benchmarks into account: when uncertainty is higher, the estimates are less likely to adhere to the benchmark. Additionally, the DABUL models produce

\noindent results with reduced shrinkage, so if the more populous second administrative areas happen to be more extreme areas, the aggregated estimate will be pulled in that direction. If estimates under the soft DABUL model do not have sufficient agreement in aggregation for a particular dataset, the practitioner may instead choose to use the hard DABUL model, with the warning that uncertainty of the benchmarks will not be accounted for. 

\subsection{Model comparison}

To compare performance of the DABUL models to the standard unit-level model in this data example, we use a leave-one-area-out cross-validation procedure. While there are cross-validation procedures for count data (\citealp{czado}), these methods are difficult to use in the context of rare events because the resulting predictive distribution is generally too noisy to meaningfully distinguish predictive ability across the narrow relevant domain. Therefore, instead of leaving out one cluster at a time, we leave out all of the clusters in one second administrative area at a time, and then predict the logit transformed design-based estimate for that second administrative area (\citealp{twocultures}). By using this approach, we avoid using the predictive distribution of a count variable, in favor of the predictive distribution of a Gaussian variable, which is much more stable. We systematically leave one second administrative area out at a time, and use each of the 3 models to predict the design-based estimate for that second administrative area. This means we can only evaluate predictive ability for second administrative areas which have valid design-based estimates and variances. This condition is met for $75\%$ of second administrative areas (87 out of 115), as the remaining $25\%$ have no observed deaths, so design-based estimates cannot be ascertained. While cross-validation is one of several ways we assess model performance, we emphasize that these results are presented with the caveat that there are valid criticisms of using cross-validation for spatially dependent data and a consensus on best practices has not yet been reached (\citealp{cv1}; \citealp{cv2}).   

We use three metrics to evaluate predictive ability. First, we calculate absolute error, which is defined as $|\hat r_{\shortminus j}- r^{D_2}_j|$, where $\hat r_{\shortminus j}$ is the second administrative level estimate for area $j$, when area $j$ data is excluded from the model, and $r^{D_2}_j$ is the design-based estimate for second administrative area $j$. Note that this metric only evaluates accuracy of the point estimate prediction with respect to the design-based point estimate. We present this standard metric with caution because it may not be the best description of model performance, as it does not take the high uncertainty of $r^{D_2}_j$ into account.  Second, we approximate the log score (or log-CPO) on the logit scale, which is defined as $\mbox{log }p_{\shortminus j}(\mbox{logit}(r^{D_2}_j))$, where $p_{\shortminus j}$ is the predictive density based on the model which excludes area $j$ data. This density does not have a closed form, but we can approximate it by 

$$p_{\shortminus j}(\mbox{logit}(r^{D_2}_j))\approx \frac{1}{B}\sum_{b=1}^B \phi(\mbox{logit}(r^{D_2}_j)|\mbox{logit}(\hat r_{\shortminus j}^{(b)}),V^{D_2}_j)$$

\noindent where $\phi(\cdot|\mbox{logit}(\hat r_{\shortminus j}^{(b)}),V^{D_2}_j)$ is the density of a normal random variable with mean, $\mbox{logit}(\hat r_{\shortminus j}^{(b)})$, and variance, $V^{D_2}_j$, and $\hat r_{\shortminus j}^{(b)}$ is the $b^{th}$ draw from the posterior distribution of second administrative level prevalence in area $j$, based on the model which excludes area $j$ data. This metric also evaluates the accuracy of the point estimate, but with respect to the whole distribution of the design-based estimate. Lastly, we calculate the interval score, a proper scoring rule (\citealp{gneiting}) which favors narrow intervals containing the true parameter value. Let $(l_{\shortminus j},u_{\shortminus j})$ denote the $90\%$ credible interval for second administrative level prevalence in area $j$, based on the model which excludes area $j$ data. Then we define the interval score for administrative area $j$ as

$$u_{\shortminus j} - l_{\shortminus j} + \frac{2}{\alpha}\mbox{max}\{0,l_{\shortminus j} - \mbox{logit}(\hat r_{\shortminus j})\}+ \frac{2}{\alpha}\mbox{max}\{0,\mbox{logit}(\hat r_{\shortminus j})-u_{\shortminus j,}\}$$ 

\noindent where we take $\alpha = 0.1$. We average each of these 3 metrics over the 87 second administrative areas and compare them across models in table \ref{tab:CVresults}. The mean absolute error is very similar across all models, with the soft DABUL model have the lowest value (0.0122). The average log score is also quite similar across models, with the standard unit-level model having the highest value (0.440). The average interval score was lowest for the soft DABUL model (3.1) and highest for the standard unit-level model (4.2). This result shows that although the DABUL intervals are more conservative, this is offset by having better coverage, so, according to this proper scoring rule, the DABUL intervals are preferred over those of the standard unit-level model.

\begin{table}
    \centering
    \caption{Summary of cross-validation metrics averaged over second administrative areas with valid design-based estimates (87 out of 115). The best scores are in \bf{bold}.}
    \begin{tabular}{c|ccc}
     & Standard unit-level (nested) & Soft DABUL &  Hard DABUL \\
    \hline
    Mean absolute error & 0.0125 & \bf{0.0122} & 0.0126 \\
    \hline
    Average log score & \bf{0.440} & 0.424 & 0.421 \\
    \hline
    Average interval score & 4.2 & \bf{3.1} & 3.9 \\
    \end{tabular}
    \label{tab:CVresults}
\end{table}

\section{\MakeUppercase{Discussion}}

We first observe that in the case of rare events and small sample size, when there are multiple nested areas, using a model with a nested structure can help reduce overshrinkage. We have demonstrated the value of a nested unit-level Bayesian model for estimating rare event prevalence which utilizes design-based estimates at a higher aggregation level. We presented two models: the hard DABUL model which imposes a hard benchmarking constraint, and the soft DABUL model which takes uncertainty of the benchmark into account, thus imposing a soft constraint. In our simulation study and application to Zambia DHS data, we observe that the DABUL models produce conservative credible intervals with much lower frequency of undercoverage, compared to the standard unit-level model. Although the hard DABUL model slightly underestimates the variance by not taking the uncertainty of the direct estimates into account, the resulting credible intervals are still fairly conservative. For this reason, and for the obvious benefit of exact agreement in aggregation, our results suggest that the hard DABUL model may be preferable over the soft DABUL model in some cases. However, our cross-validation study suggests, via average interval scores, that while both DABUL models provide credible intervals which are superior to that of the standard unit-level model, the soft DABUL model provides the credible intervals with the best performance. Further study is necessary to determine the costs and benefits of choosing between the two DABUL models, however, the soft DABUL model remains the more conservative, principled approach because of its proper accounting for the large uncertainty of the benchmarks. 

One criticism of the use of generalized linear unit-level models for complex survey data, is that they do not directly take sample design into account. Solutions have been developed for the case of linear models (\citealp{YouRao2002}), but they do not extend to the generalized linear case (\citealp{twocultures}).  While the DABUL models do not completely solve this open problem, they are an improvement in the sense that they take design-based estimates at a higher aggregation level into account.

The DABUL models could also be extended to include covariates, which is encouraged whenever covariates with predictive power are available. We do not include covariates here because previous work has shown that available covariates in LMICs have little effect on predictive power in estimating NMR when spatial random effects are already included in the model (\citealp{golding},\citealp{wakefieldCovariates}). It is straightforward to include covariates in the DABUL models, though the aggregation step can introduce additional error.

Because the DABUL models, as derived, use a Negative Binomial sampling model, they require the outcome to be a rare event. While this model could theoretically be extended to estimate prevalence of non-rare events by using a Binomial or Beta-Binomial sampling model, these distributions do not enjoy the same additive properties as the Poisson and Negative Binomial distributions. As a result, the distributions of $Y_{i+}$ and $\tilde Y_i|Y_{i+}$ quickly become unwieldy as they require summing over a complete enumeration of the state space.  Conversely, this model could easily be extended to continuous outcomes because the Normal distribution does share these additive properties. The only necessarily changes to the model would be to replace (\ref{eq:DCM1}) and (\ref{eq:Ypluslikelihood}) with the normal distributions implied by the Gaussian likelihood.

\newpage

\noindent \textbf{{\large Data Availability Statement}}

\noindent The processed 2014 Zambia DHS data used in this article are available in this article's online supplementary material.

\noindent \textbf{{\large Declaration of Interest}}

\noindent We declare no conflicts of interest.

\noindent This work was supported by the National Institutes of Health [R01 HD112421-02].

\bibliography{main}

\newpage

\noindent {\bf Figure 1.} Summary of births and neonatal deaths between 2009 and 2013
recorded in the 2014 Zambia DHS.

\noindent {\bf Figure 2.} Map of NMR estimates in Zambia (2009-2013) under various models.

\noindent {\bf Figure 3.} NMR estimates at the second administrative level in Zambia (2009-
2013) under various models.

\noindent {\bf Figure 4.} Comparison of DABUL models with the standard unit-level Bayesian
model.

\noindent {\bf Figure 5.} Second administrative area neighborhood structure for simulation
study

\noindent {\bf Figure 6.} Discrepancy between aggregated model-based second administrative
level estimates and design-based first administrative level estimates, across 500
simulations for each of 4 settings.

\noindent {\bf Figure 7.} Absolute error of second administrative level estimates, across 500 simulations for each of 4 settings.

\noindent {\bf Figure 8.} Coverage of 90\% credible intervals for second administrative level estimates, across 500 simulations for each of 4 settings.

\noindent {\bf Figure 9.} Width of 90\% credible intervals (in deaths per 1000 births) for second
administrative level estimates, across 500 simulations for each of 4 settings.

\noindent {\bf Figure 10.} Map of NMR estimates at the second administrative level in Zambia
(2009-2013) using the DABUL models, as compared to a standard unit-level
Bayesian model, both with BYM2 spatial effects at the second administrative
level nested based on first administrative level.

\noindent {\bf Figure 11.} Width of 90\% credible intervals of NMR estimates (in deaths per 1000 live births) at the second administrative level in Zambia (2009-2013) using
the DABUL models, as compared to a standard unit-level Bayesian model.

\noindent {\bf Figure 12.} NMR estimates at the second administrative level in Zambia (2009-
2013) using the soft DABUL model, as compared to a standard unit-level nested
Bayesian model.

\newpage

\appendix

\begin{center}
\LARGE
    Direct-Assisted Bayesian Unit-level Modeling for Small Area Estimation of Rare Event Prevalence
\end{center}

\hspace{10pt}

\large
\noindent Alana McGovern$^{1*}$, Katherine Wilson$^2$, and Jon Wakefield$^{1,2}$

\hspace{8pt}

\small  
\noindent$^1$ Department of Statistics, University of Washington, Seattle WA, USA\\
$^2$ Department of Biostatistics, University of Washington, Seattle WA, USA

\hspace{8pt}
\normalsize

\noindent $^*$Corresponding author, amcgov@uw.edu

\newpage
\renewcommand{\thesection}{{Appendix} \Alph{section}}
\renewcommand{\thefigure}{S\arabic{figure}}
\setcounter{figure}{0}

\section{Definitions and Notation \label{notation appendix}}

$\odot$ denotes the Hadamard product, i.e., for 2 $m$-length vectors, ${\bf x}$ and ${\bf y}$, ${\bf x}\odot{\bf y}$ is an $m$-length vector with elements $({\bf x}\odot{\bf y})_i={\bf x}_i{\bf y}_i$.

\begin{table}[h]
    \begin{tabular}{cc||c}
    {\bf Element} & {\bf Description} &  \\
    \hline\hline
    $m_k$ & number of $k^{th}$ administrative areas & \\
    $\Delta$ & set containing all clusters & \\
    $\delta_k(j)$ & set containing all clusters in $k^{th}$ admin area $j$ & \\
    $(\delta_U,\delta_R)$ & set of all urban and rural clusters, respectively & \\
    \hline\hline
        {\bf Element} & {\bf Description} & {\bf Corresponding vector(s)} \\
       \hline\hline
        $\gamma_c$ & indicator variable denoting whether & $\boldsymbol{\gamma}:=\{\gamma_c:c\in\Delta\}$ \\
         & cluster $c$ was sampled & $\boldsymbol{\gamma}_i:=\{\gamma_c:c\in\delta_1(i)\}$ \\
         \hline
        $n_c$ & \shortstack{number of births in the sampled \\ households in cluster $c$} & ${\bf n}:=\{n_c:c\in\Delta\}$ \\ 
        & (equals $0$ if cluster $c$ was not sampled) & \\
        \hline
         & number of births in all & ${\bf N}:=\{N_c:c\in\Delta\}$ \\
        $N_c$ & households in cluster $c$ & ${\bf N}_i:=\{N_c:c\in\delta_1(i)\}$ \\
         &  & ${\bf N}^{(s)}_i:=\{N_c:\gamma_c=1\}$ \\
         \hline
        $N_{i+}:=\sum\limits_{c\in\delta_1(i)}N_c$ & \shortstack{number of births in \\ $1^{st}$ administrative area $i$} & ${\bf N}_+:=\{N_{i+}:i=1,...,m_1\}$ \\
        \hline
        $Z_c$ & \shortstack{number of neonatal deaths in the sampled \\ households in cluster $c$ } & ${\bf Z}:=\{Z_c:c\in\Delta\}$ \\
         & (equals $0$ if cluster $c$ was not sampled) & ${\bf Z}_i:=\{Z_c:c\in\delta_1(i)\}$ \\
        \hline
         & number of neonatal deaths in all & ${\bf Y}:=\{Y_c:c\in\Delta\}$ \\
        $Y_c$ & households in cluster $c$ & ${\bf Y}_i:=\{Y_c:c\in\delta_1(i)\}$ \\
        &  & ${\bf Y}^{(s)}:=\{Y_c:\gamma_c=1\}$ \\
         &  & ${\bf Y}^{(s)}_i := {\bf Y}_i\cap {\bf Y}^{(s)}$\\
         \hline
        $Y_{i+}:=\sum\limits_{c\in\delta_1(i)}Y_c$ & \shortstack{number of neonatal deaths in \\ $1^{st}$ administrative area $i$} & ${\bf Y_+}:=\{Y_{i+}:i=1,...,m_1\}$ \\
        \hline
        $Y_{i+}^{(s)}:=\sum\limits_{\substack{c\in\delta_1(i)\\ \gamma_c=1}}Y_c$ & \shortstack{number of neonatal deaths in the sampled \\ households in $1^{st}$ administrative area $i$} & ${\bf Y^{(s)}_+}:=\{Y^{(s)}_{i+}:i=1,...,m_1\}$ \\
        \hline
        $r_c$ & risk of neonatal death in cluster $c$ & ${\bf r}:=\{r_c:c\in\Delta\}$ \\
         &  & ${\bf r}_i:=\{r_c:c\in\delta_1(i)\}$ \\
         \hline
         $r^{D_k}_i$ & design-based prevalence estimate & ${\bf r^{D_k}}:=\{r^{D_k}_i:i=1,...,m_k\}$ \\
          & for $k^{th}$ administrative area $i$ & \\
         \hline
         $V^{D_k}_j$ & design-based variance of logit$(\hat r^{D_k}_i)$ & ${\bf V^{D_k}}:=\{V^{D_k}_j:i=1,...,m_k$\} \\
         \hline

    \end{tabular}
    \label{tab:notation}
\end{table}

\newpage

\section{Comparison of spatial models for Zambia NMR \label{spatial_models}}
We compare NMR estimates and uncertainty under model (\ref{eq:stratnested}) using IID and BYM2 spatial effects. From these figures, we observe that, in this example, choice of spatial effect model makes little difference, so we choose BYM2 as it provides the option of more structure (by including an ICAR component in addition to an IID component) and has the added benefit of allowing more informative estimation in areas without observed data. Note that uncertainty of estimates under the BYM2 spatial effect is lower on average, and those for which it is higher are areas which have both low sample size and surrounding neighbors with low sample size.

\begin{figure}[h]
    \centering
    \caption{Second administrative level NMR estimates}
    \includegraphics[width=0.8\linewidth]{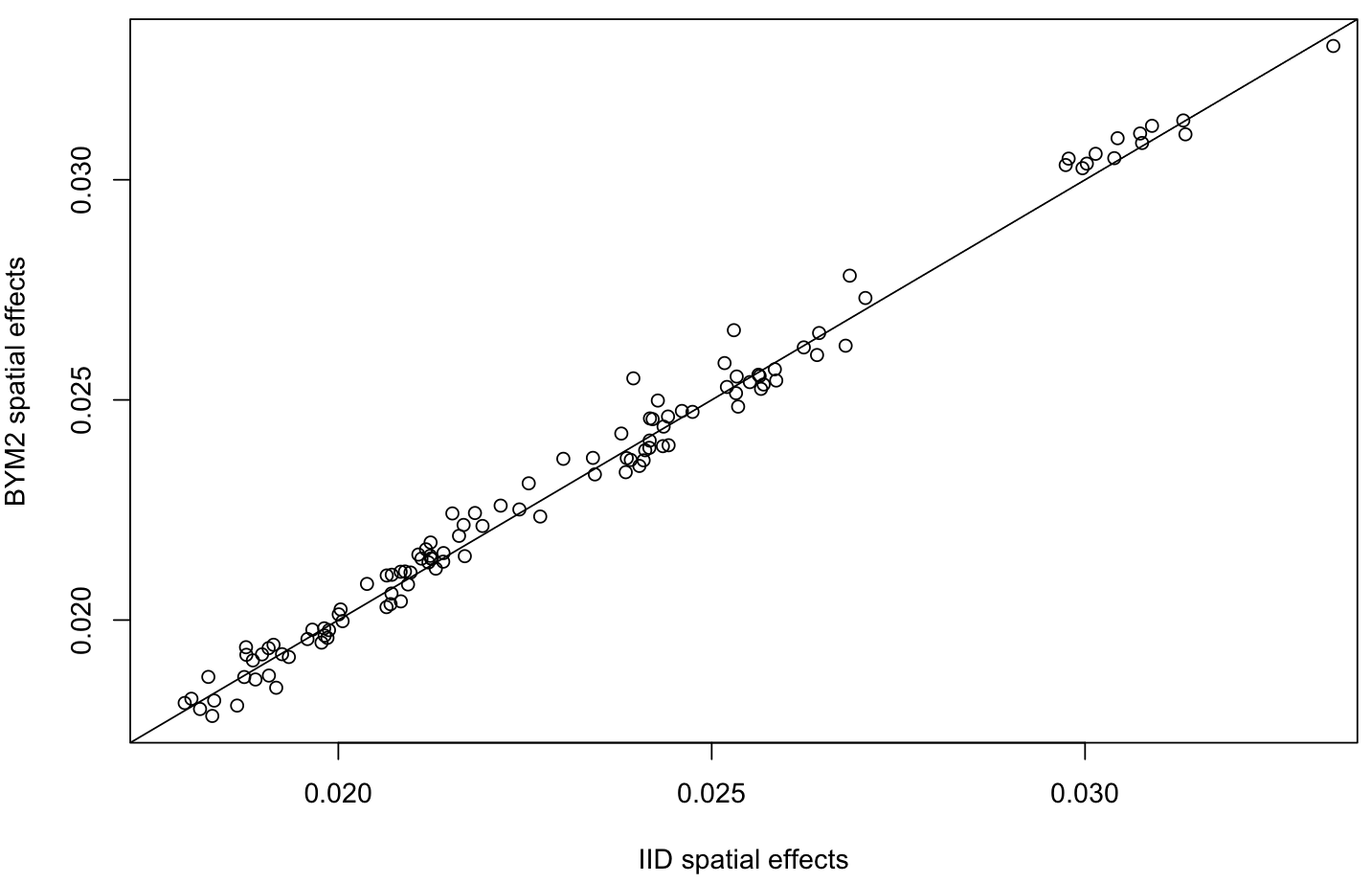}
    \label{fig:compare_spatial_est}
\end{figure}
\newpage

\begin{figure}[h]
    \centering
    \caption{Width of 90\% credible intervals for second administrative level NMR estimates}
    \includegraphics[width=0.8\linewidth]{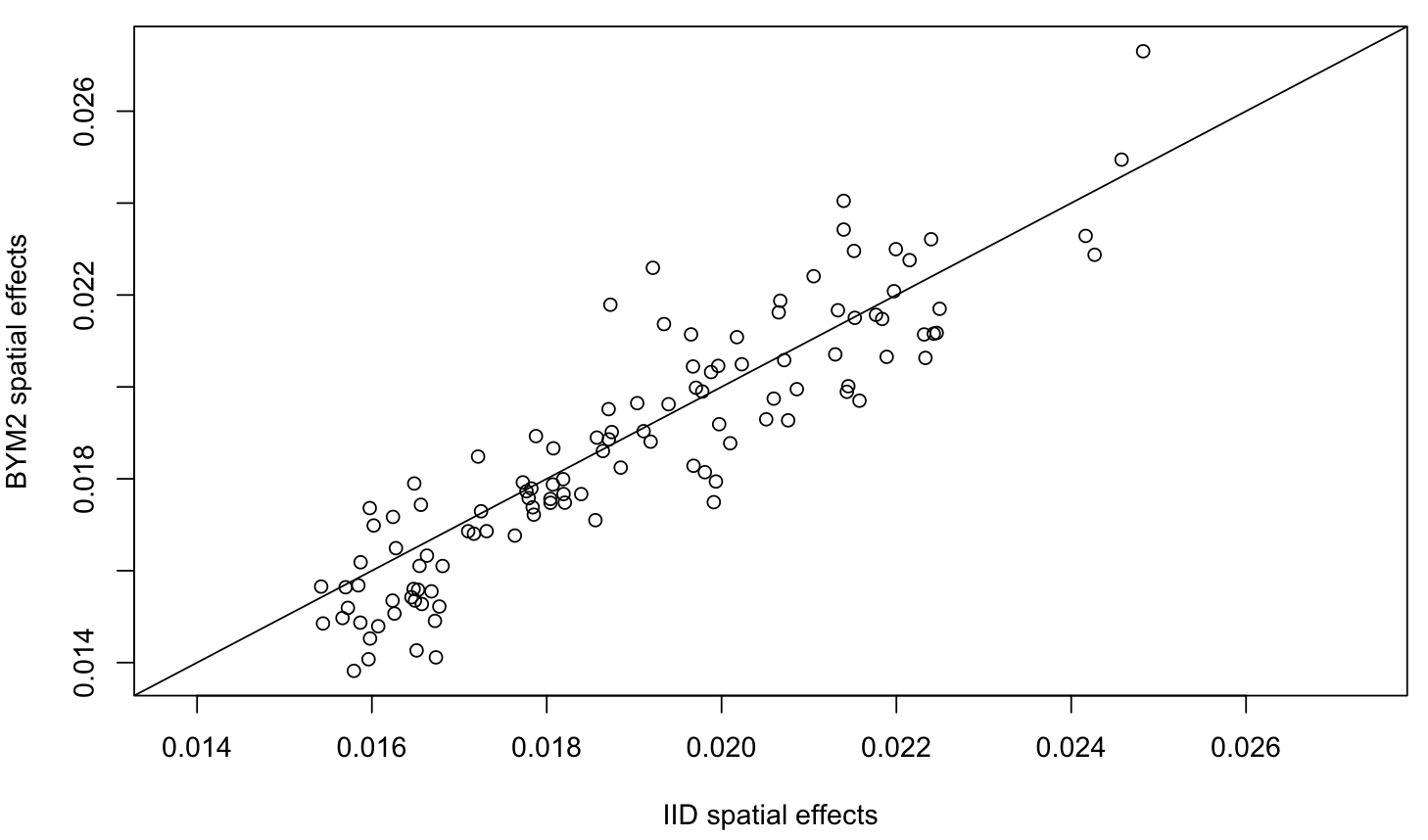}
    \label{fig:compare_spatial_uncertainty}
\end{figure}
\newpage

\section{Priors on regression and overdispersion parameters}\label{Priors Appendix}

\begin{figure}[h]
    \centering
    \caption{Exponential transformations of Normal distributions with mean $-3.5$ and variances $1^2$, $3^2$, and $10^2$, respectively.}
    \includegraphics[width=1\linewidth]{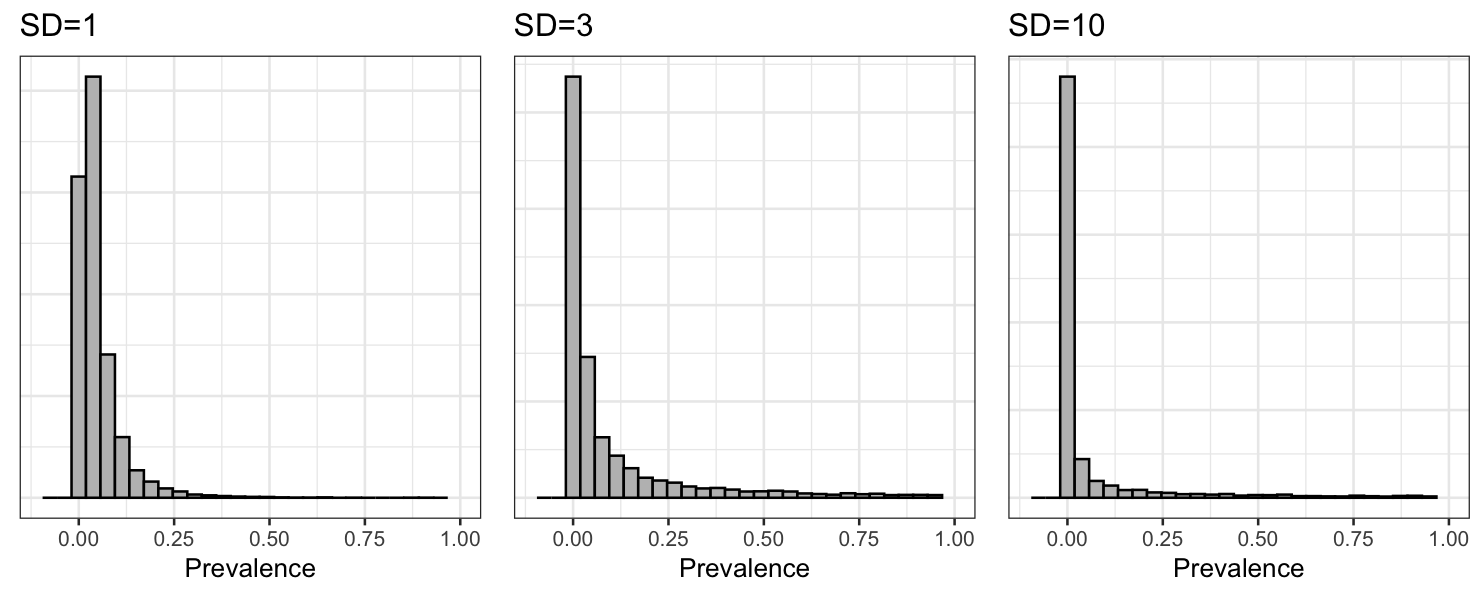}
    \label{fig:expnormals}
\end{figure}

The priors on the regression parameters $(\alpha_U,\alpha_R,\boldsymbol{\beta})$ are chosen with the goal of being relatively diffuse on the exponential scale, while still accounting for the prior belief that the outcome measure is a rare event. Figure \ref{fig:expnormals} displays how variability of the Normal distribution is translated on the exponential scale in the case of rare events. Note that when the variance on the linear scale is large, as in the panel on the right, the distribution on the exponential scale has very high probability near 0, so the variability on the desired scale actually \textit{decreases}.  From this figure we can deduce that the chosen prior on $(\alpha_U,\alpha_R)$, $\mathcal{N}(-3.5,3^2)$, is in fact a diffuse prior in this context. The chosen prior for the fixed effects $\boldsymbol{\beta}$, $\mathcal{N}(0,1)$, has a smaller variance to avoid a prior on the linear predictor with such high variability that the implied prior on the exponential scale shrinks towards 0.

\begin{figure}[h]
    \centering
    \caption{Posterior distribution of the overdispersion parameter in the nested negative binomial model, fit to Zambia neonatal mortality data, under two different choices of prior.}
    \includegraphics[width=0.75\linewidth]{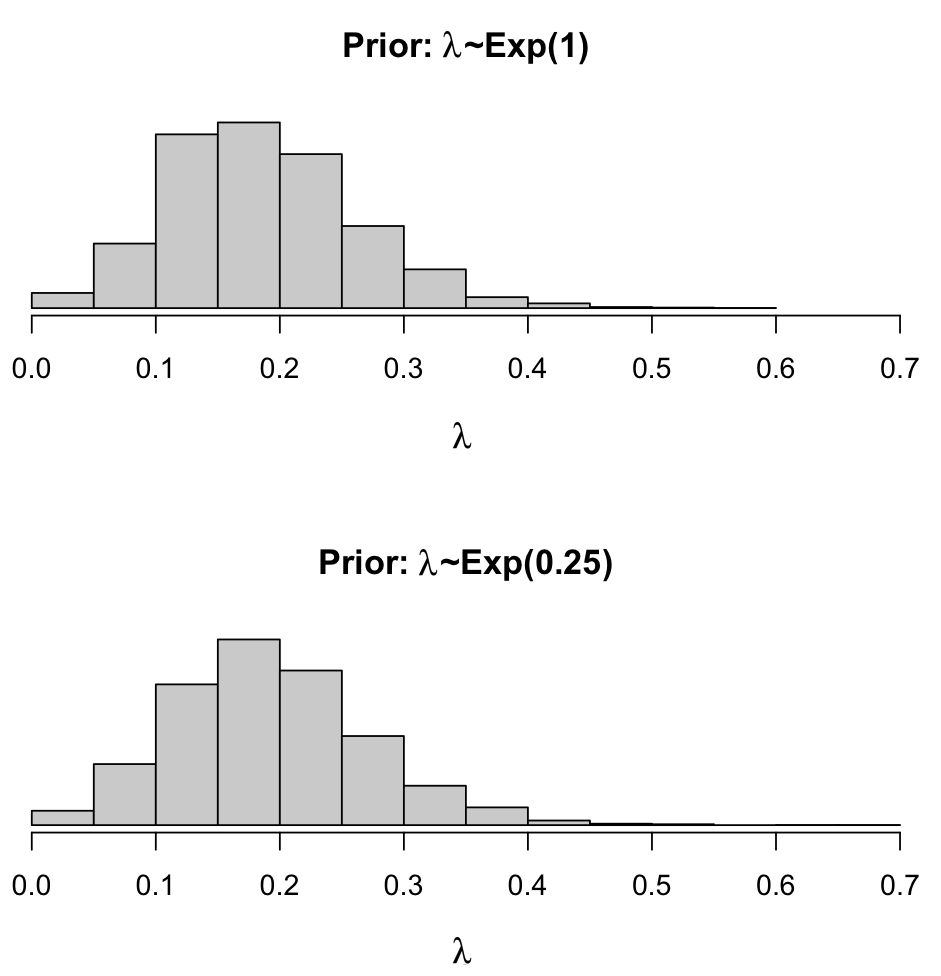}
    \label{fig:overdisp_post}
\end{figure}

The prior on the overdispersion parameter, Exp$(1)$, is chosen to reflect the belief that the variance of the outcome is likely to be less than two times its mean (with $63\%$ probability), and that its more than four times the mean with fairly low probability, $(5\%)$. This is a reasonable assumption, because if the overdispersion is much larger than $1$, the increased variability implies that the outcome is not rare in some areas. To test the sensitivity of this prior choice we fit regression model (\ref{eq:stratnested}) to the Zambia DHS data using the chosen prior and using an alternative prior, Exp$(0.25)$, which puts much less weight near $0$. Figure \ref{fig:overdisp_post} shows that the resulting posterior distributions of the overdispersion parameter are nearly identical.

\newpage
\section{Coefficient of variation for Zambia NMR estimates in motivating example \label{CV}}

\begin{figure}[h!]
    \centering
    \caption{The maps on the top row display coefficients of variation for the NMR estimates under unit-level Bayesian negative binomial models with BYM2 spatial effects at the second administrative level, while the maps on the bottom row display coefficients of variation for the H{\'a}jek direct estimates at the first and second administrative level, respectively.  The range of coefficients of variation for the non-nested and nested model are $10-16\%$ and $18-32\%$, respectively, while the range of coefficients of variation for the first and second administrative level direct estimates are $18-32\%$ and $13-173\%$, respectively. The latter reflects the extremely high uncertainty of the second administrative level direct estimates, even among those areas which have sufficient data to make valid estimates.}
    \includegraphics[width=0.95\linewidth]{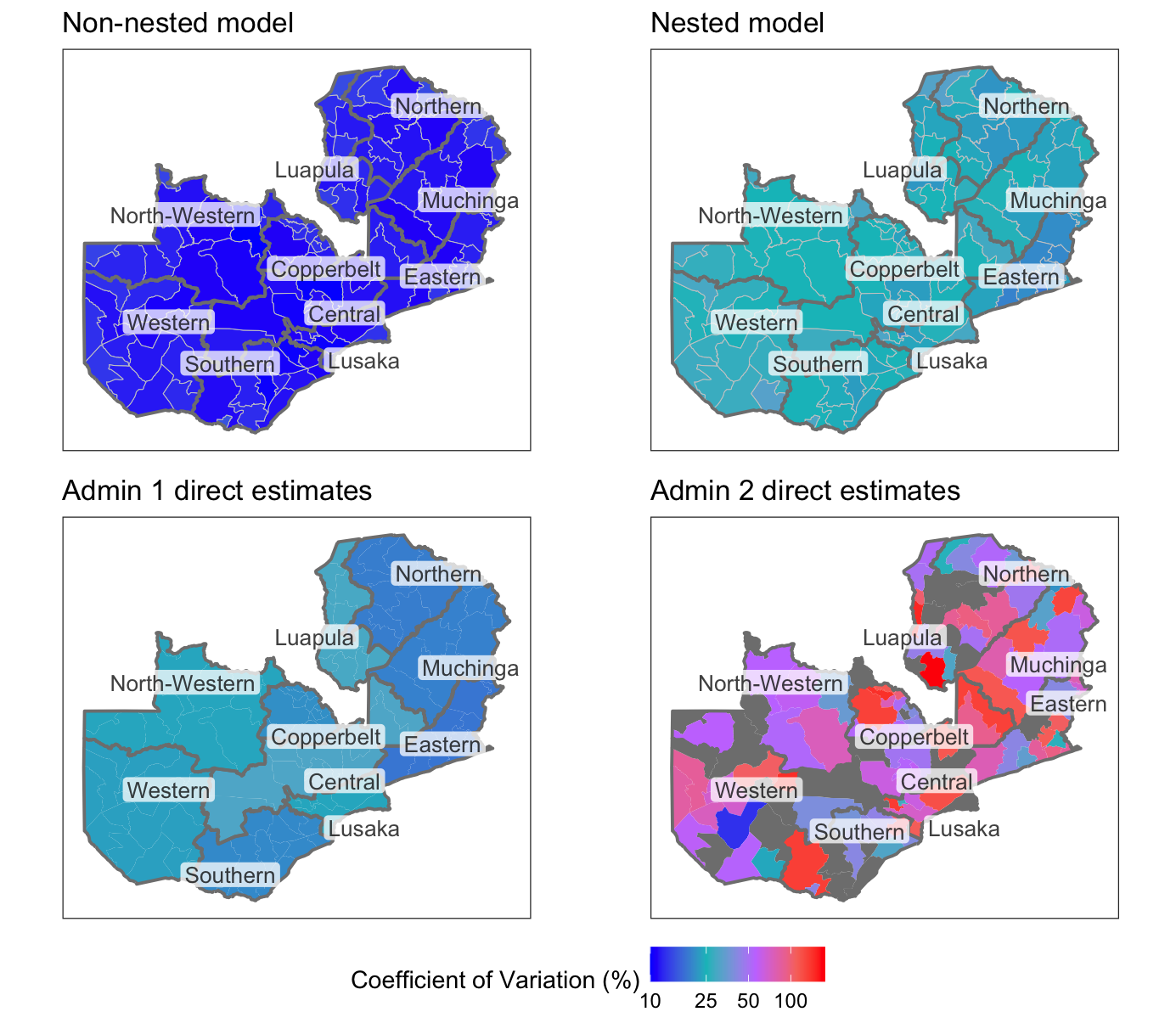}
\end{figure}

\newpage

\section{Additional simulation results \label{Appendix A}}

\begin{figure}[h]
\caption{Discrepancy between aggregated model-based second administrative level estimates and design-based first administrative level estimates, among the subset of simulations for which the aggregated standard unit-level model estimate has a discrepancy with the direct estimate that is larger than $0.001$.}
    \centering
    \includegraphics[scale=0.28]{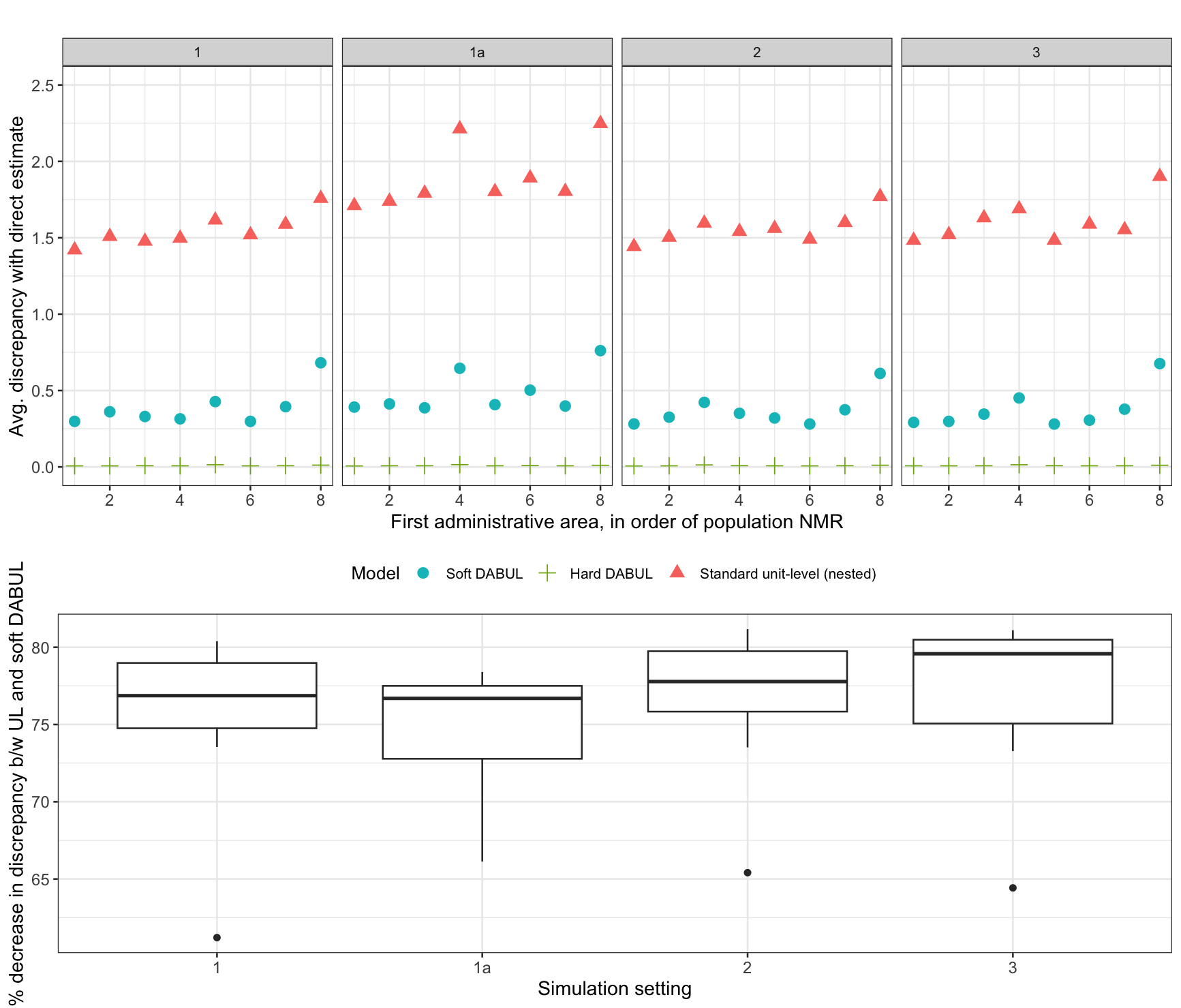}
    \label{fig:SimDisc2}
\end{figure}

Figure \ref{fig:SimDisc2} is similar to figure \ref{fig:SimDisc1}, except we only examine the areas for which the standard unit-level model has significant discrepancy with the direct estimates (greater than 0.001). We observe similar patterns to those in the previous figure, but the decrease in discrepancy, displayed in the bottom panel, is more significant, with aggregated soft DABUL estimates having  $74.7$-$76.8\%$ lower discrepancy with the first administrative level direct estimates, on average, compared to aggregated estimates from the standard unit-level model. In other words, when there is larger discrepancy between the aggregated standard unit-level estimates and direct estimates, the soft DABUL model reduces the discrepancy with direct estimates by a larger magnitude. 

\begin{figure}[h]
\caption{{\bf Coefficient of variation for second administrative level estimates, across 500 simulations for each of 4 settings.}}
    \centering
    \includegraphics[scale=0.35]{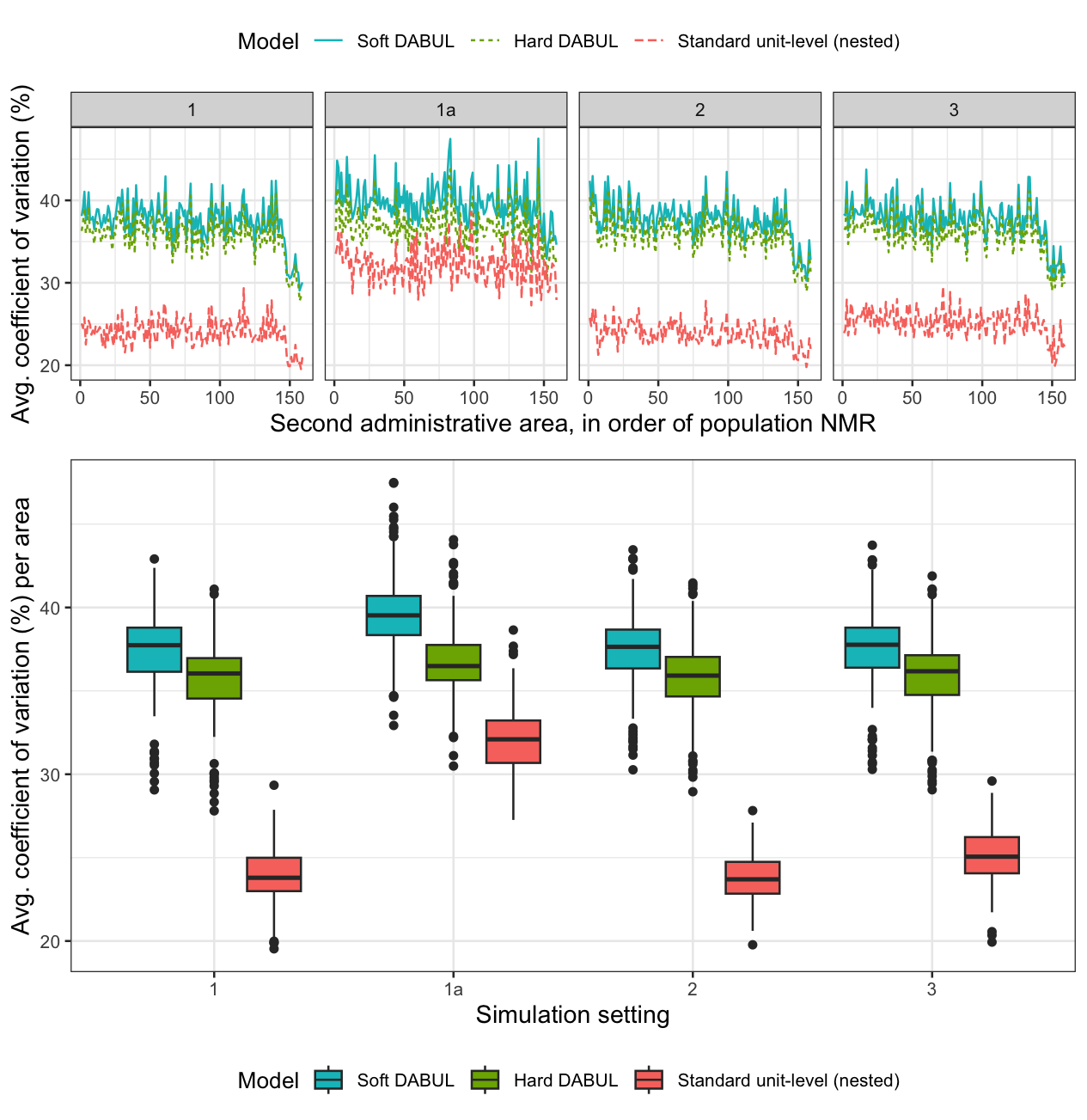}
    \label{fig:SimCoefVar}
\end{figure}

\newpage

\section{Additional plot of Zambia NMR estimates \label{Appendix B}}

\begin{figure}[!h]
    \centering
    \caption{NMR estimates (deaths per 1000 live births) at the second administrative level in Zambia (2009-2013) using the DABUL models, as compared to a standard unit-level nested Bayesian model, both with BYM2 spatial effects at the second administrative level nested based on first administrative level.}
    \includegraphics[scale=0.28]{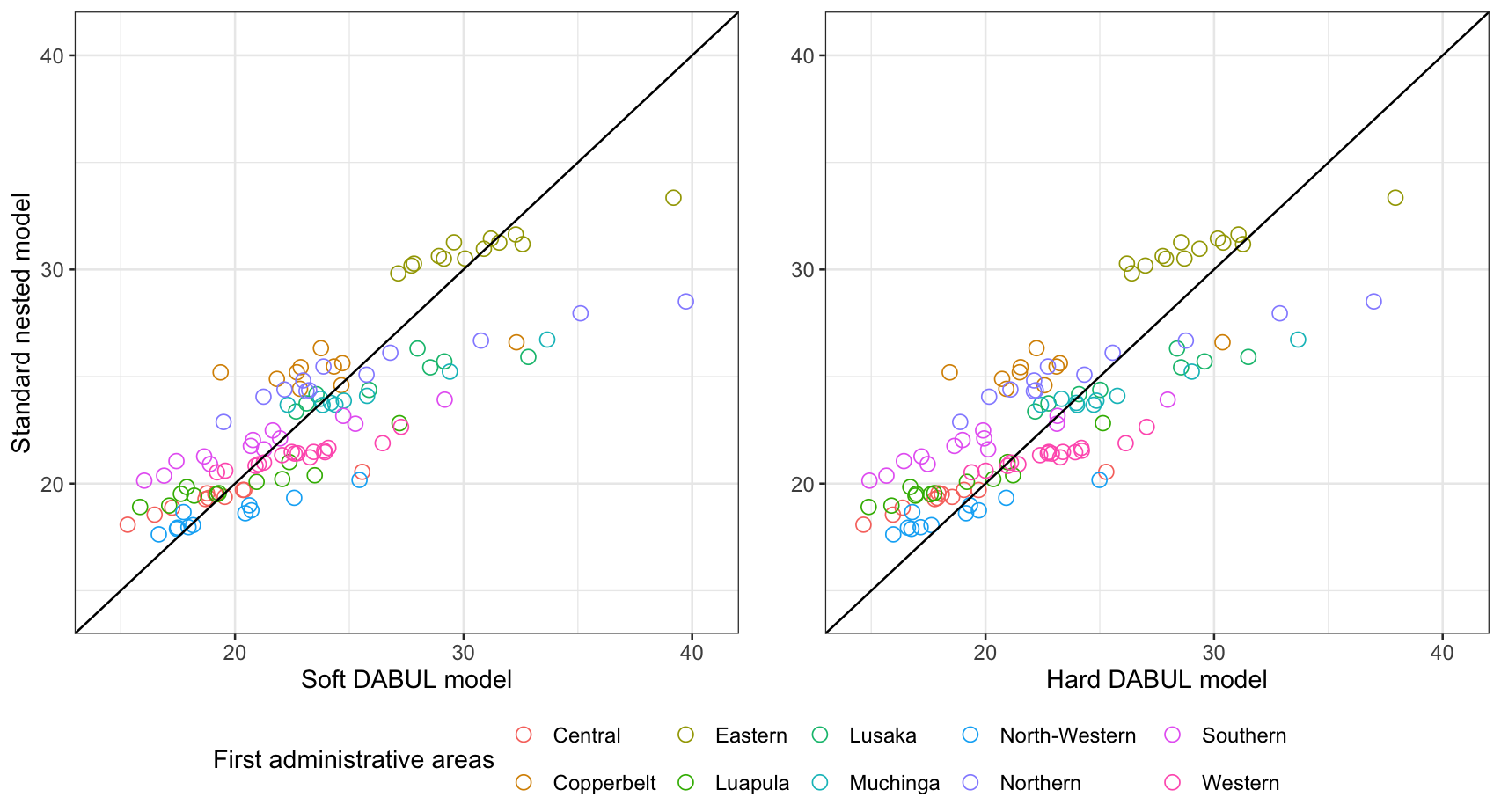}
    \label{fig:DABULscatter}
\end{figure}

\end{document}